\documentstyle[preprint,prl,aps,epsfig]{revtex}
\tightenlines
\begin{document}

\title{Heavy Quark Photoproduction in Ultra-peripheral Heavy Ion Collisions}

\author{Spencer R. Klein$^1$, Joakim Nystrand$^2$, and Ramona Vogt$^{1,3}$} 
\address{$^1$Lawrence Berkeley National Laboratory, Berkeley, CA 94720 \break
$^2$Department of Physics, Lund University, Lund SE-22100, Sweden \break
$^3$Physics Department, University of California, Davis, CA 95616} 

\break 
\maketitle
\vskip -.2 in
\begin{abstract}
\vskip -.2 in 

Heavy quarks are copiously produced in ultra-peripheral heavy ion
collisions.  In the strong electromagnetic fields, $c\overline c$ and
$b\overline b$ are produced by photonuclear and two-photon
interactions.  Hadroproduction can also occur in grazing interactions.  We
calculate the total cross sections and the quark transverse momentum and
rapidity distributions, as well as the $Q\overline Q$ invariant mass
spectra from the three production channels.  We consider $AA$ and $pA$
collisions at the Relativistic Heavy Ion Collider and the Large Hadron
Collider.  We discuss techniques for separating the three processes
and describe how the $AA$ to $pA$ production ratios might be measured
accurately enough to study nuclear shadowing.

\end{abstract}

\section{Introduction}
\label{introsec}

In ultra-peripheral heavy ion collisions, heavy quarks can be produced
in electromagnetic or hadronic interactions.  Electromagnetic
production occurs through strong electromagnetic fields which can
interact with a target nucleus in the opposing beam (photoproduction)
or with the electromagnetic field of the opposing beam (two-photon
reactions) and produce hadronic final states, including heavy quark
pairs.  Hadroproduction of heavy quark pairs is also possible in
grazing interactions.  Since photon emission is coherent over the
entire nucleus and because the photon is colorless, the three channels
can be distinguished by the presence of zero, one or two rapidity gaps
in the events and by whether or not the nuclei dissociate.

Many types of ultra-peripheral collisions have been
considered\cite{baurrev}.  Photonuclear interaction studies have
included Coulomb dissociation\cite{ZDCanalysis} and coherent vector
meson production\cite{usPRC,parkcity}.  Final states studied in
two-photon interactions have included production of lepton
pairs\cite{INPC}, single mesons and meson pairs\cite{FELIX} as well as
production of the Higgs boson and other exotica \cite{kp}. Although
the list of experimentally observed channels is currently short, as
RHIC gears up new results should come quickly and measurements of
heavy quark production in ultra-peripheral collisions may not be too
far off.
 
We build on previous calculations of heavy quark
photoproduction\cite{bbar,bbar2,knv1} and two-photon
production\cite{vidovictwophoton} in heavy ion collisions.  For the
first time, we consider resolved photon processes. We use modern
parton distribution functions along with up-to-date accelerator
species and luminosities.  We also consider hadroproduction in grazing
collisions and compare the three channels.  Finally, since useful
measurements of shadowing will require high accuracy, we consider the
uncertainties inherent in these calculations and methods to control
them.

We will compare the total cross sections, heavy quark transverse momentum,
$p_T$, and rapidity, $y$, distributions, and the $Q
\overline Q$ pair invariant mass, $M$, distributions
in all three channels.
We also discuss $pA$ collisions to see how a comparison of
photoproduction in $AA$ ($\gamma A$) and $pA$ (effectively $\gamma
p$) can be used to study nuclear effects on the parton distribution
functions (shadowing).  

For consistency, we will use parallel approaches to all three
calculations with the same quark mass, parton distribution
and QCD scale.  For the charm quark mass, $m_c$, this is somewhat
problematic because studies of different production channels seem to
prefer different values.  Hadroproduction calculations have typically
best fit the data with a relatively light $m_c$, $\sim 1.2-1.4$ GeV
\cite{hpc}, while photoproduction studies have generally favored
higher values, $1.5-1.8$ GeV\cite{frixione,berezhnoy}.  The limited
two-photon data favors an intermediate value, around 1.6
GeV\cite{Drees,l3}.  For the bottom quark, the typical mass range
considered is $4.5-5.0$ GeV.  Hadroproduction calculations with
$m_b=4.75$ GeV underpredict the $b\overline b$ cross section observed
in 1.8 TeV $p\overline p$ collisions\cite{CDF}, suggesting that a
smaller mass might be preferred \cite{CDF}.  Some exotic $b$
production mechanisms have been suggested to explain the excess
\cite{harris,other}.  Lattice studies suggest lower $c$ and $b$ quark
masses, $1.1-1.4$ GeV for charm and $4.1-4.4$ GeV for bottom (the pole
masses are somewhat higher) \cite{PDG}.  We use $m_c=1.2$ GeV and
$m_b=4.75$ GeV throughout this paper but we will also discuss the
effect of varying the mass and scale. The QCD scale entering the
running coupling constant, $\alpha_s(Q^2)$, and the parton distribution
functions are proportional to the transverse masses, $Q^2 = 4m_T^2$ for
charm and $m_T^2$ for bottom \cite{hpc}.

Table~\ref{lum} gives recent estimates of nucleon-nucleon center of
mass energies, $\sqrt{S}$, and luminosities for $AA$ and $pA$
collisions at RHIC\cite{RHIClum} and the
LHC\cite{brandt,brandt2}.  The LHC luminosities assume that two
experiments take data and that there is a 125 ns bunch spacing.  
At RHIC, the $pA$
energies are the same as the $AA$ energies while the luminosities are
taken to be the geometric mean of the $AA$ and $pp$ luminosities
(${\cal L}_{pp} = 1.4 \times 10^{31}$ cm$^{-2}$ s$^{-1}$).  There is
also the possibility of $pA$ collisions at the LHC.  However, at the
LHC, the proton and the ion must have the same magnetic rigidity.
Because protons and ions have different charge to mass
ratios, they have different per nucleon energies and the
center of mass is no longer at rest in the lab.  These $pA$ collisions
are then at somewhat higher per nucleon energies than the corresponding
$AA$ collisions.  At the LHC, the $pA$ collision rates can exceed 200
kHz although some experiments may need to run at a lower luminosity.  Because
RHIC is a dedicated heavy ion accelerator, it is likely to run a wider
variety of beams than the LHC.  At RHIC, $dA$ collisions may be an
alternative or supplement to $pA$.  Except for the different initial
isospin, most of the $pA$ discussion should also hold for $dA$ because
shadowing is small in deuterium.

Section \ref{photosec} discusses photoproduction of heavy quarks in
$AA$ and $pA$ collisions.  Section \ref{hadrosec} covers
hadroproduction in peripheral $AA$ and minimum bias $pA$ collisions.
Section \ref{gamgamsec} considers $\gamma \gamma \rightarrow Q
\overline Q$ production.  Section \ref{thsec} compares our results for
the three channels while section \ref{exsec} is dedicated to a
discussion of how to disentangle the production channels
experimentally.  In section \ref{sumsec}, we draw our conclusions.

\section{Photoproduction}
\label{photosec}

\begin{figure}
\setlength{\epsfxsize=0.95\textwidth}
\setlength{\epsfysize=0.5\textheight}
\centerline{\epsffile{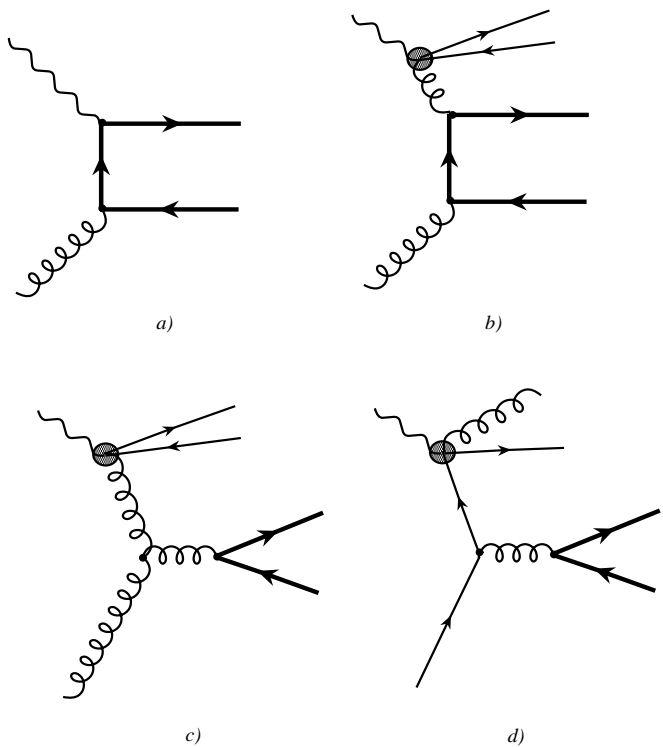}}
\caption[]{Feynman diagrams for heavy quark photoproduction for (a) direct
and (b)-(d) resolved photons.  The crossed diagrams for (a) and (b) are
not shown.}
\label{Feynphot}
\end{figure}

Photoproduction of heavy quarks occurs when a photon emitted from one
nucleus fuses with a gluon from the other nucleus, forming a
$Q\overline Q$ pair\cite{bbar,bbar2,bertulani} (``direct''
production), as in Fig.~\ref{Feynphot}(a). 
The photon can also fluctuate into a state with multiple $q\overline
q$ pairs and gluons, {\it i.e.}\ $|n(q \overline q)m(g)\rangle$. One
of these photon components can interact with a quark or gluon from the
target nucleus (``resolved'' production), as in
Figs.~\ref{Feynphot}(b)-(d)\cite{witten}.  The photon components are
described by parton densities similar to those used for protons except
that no useful momentum sum rule applies to the photon \cite{jpgconf}.

At leading order (LO), the partonic cross section of the direct
contribution is proportional to $\alpha \alpha_s(Q^2) e_Q^2$, where
$\alpha_s(Q^2)$ is the strong coupling constant, $\alpha=e^2/\hbar c$ is the
electromagnetic coupling constant, and $e_Q$ is the quark charge, $e_c = 2/3$
and $e_b = -1/3$.  The
resolved partonic cross section is proportional to $\alpha_s^2(Q^2)$.  Even
though the resolved partonic cross sections are
larger than the direct partonic cross section, the smaller flux of quarks and
gluons from the photon suggests that the resolved contribution 
should be smaller than the direct component.

The cross sections are calculated using the Weizs\"acker-Williams
virtual photon flux, modern parameterizations of the target gluon and
quark distributions and the LO partonic cross sections.  Newer parton
distributions are considerably softer than the flat, scaling
parameterizations used earlier\cite{bbar,bbar2}. We require that the
photoproduction not be accompanied by hadronic
interactions\cite{bbar,bbar2}.  This could be done by restricting the
impact parameter, $b$, to greater than twice the nuclear radius,
$R_A$.  Here we weight the $b$-dependent photoproduction probability
by the $b$-dependent hadronic non-interaction probability.

Direct $Q\overline Q$ pairs are produced in the reaction $\gamma(k) +
N(P_2) \rightarrow Q(p_1) + \overline Q(p_2) + X$ where $k$ is the
four momentum of the photon emitted from the virtual photon field of
the projectile nucleus, $P_2$ is the four momentum of the interacting
nucleon $N$ in ion $A$, and $p_1$ and $p_2$ are the four momenta of
the produced $Q$ and $ \overline Q$.  The photons are almost
real.  Their slight virtuality, $|q^2| < (\hbar c/R_A)^2$,
is neglected.

On the parton level, the photon-gluon fusion reaction is $\gamma(k) +
g(x_2 P_2) \rightarrow Q(p_1) + \overline Q(p_2)$ where $x_2$ is the
fraction of the target momentum carried by the gluon.  The LO $Q
\overline Q$ photoproduction cross section for quarks with mass $m_Q$
is \cite{joneswyld}
\begin{equation}
s^2 \frac{d^2 \sigma_{\gamma g}}{dt_1 du_1} = \pi \alpha_s(Q^2) \alpha e_Q^2 
B_{\rm QED} (s,t_1,u_1) \delta(s + t_1 + u_1)
\label{gamApart}
\end{equation}
where
\begin{equation}
B_{\rm QED} (s,t_1,u_1) = 
\frac{t_1}{u_1} + \frac{u_1}{t_1} + \frac{4m_Q^2 s}{t_1 u_1} \left[ 1 - 
\frac{m_Q^2 s}{t_1 u_1} \right] \, \, .
\label{bqeddef}
\end{equation}
Here $\alpha_s(Q^2)$ is evaluated to one loop at scale $Q^2$.  The
partonic invariants, $s$, $t_1$, and $u_1$, are defined as $s = (k +
x_2 P_2)^2$, $t_1 = (k - p_1)^2 - m_Q^2 = (x_2 P_2 - p_2)^2 - m_Q^2$,
and $u_1 = (x_2 P_2 - p_1)^2 - m_Q^2 = (k - p_2)^2 - m_Q^2$.  In this
case, $s = 4k \gamma_L x_2m_p$ where $\gamma_L$ is the Lorentz boost
of a single beam and $m_p$ is the proton mass.  Since $k$ ranges over
a continuum of energies up to $E_{\rm beam} = \gamma_L m_p$, we define
$x_1 = k/P_1$ analogous to the parton momentum fraction where $P_1$ is
the nucleon four momentum. For a detected quark in a nucleon-nucleon
collision, the hadronic invariants are then $S = (P_1 + P_2)^2$, $T_1
= (P_2 - p_1)^2 - m_Q^2$, and $U_1 = (P_1 - p_1)^2 - m_Q^2$.

We label the quark rapidity as $y_1$ and the antiquark rapidity as $y_2$.
The quark rapidity is related to the invariant $T_1$ by $T_1 = -
\sqrt{S} m_T e^{-y_1}$ where $m_T = \sqrt{p_T^2 + m_Q^2}$.  The
invariant mass of the pair can be determined if both the $Q$ and
$\overline Q$ are detected.  The square of the invariant mass, $M^2 =
s = 2m_T^2 (1 + \cosh(y_1 - y_2))$, is the partonic center of mass
energy squared.  The minimum photon momentum necessary to produce a $Q
\overline Q$ pair is $k_{\rm min} = M^2/4\gamma_L m_p$.  At LO, $x_2
= (m_T/\sqrt{S})(e^{y_1} + e^{y_2})$ and $x_1 =
(m_T/\sqrt{S})(e^{-y_1} + e^{-y_2})$.  We calculate $x_1$ and $x_2$ as
in an $NN$ collision and then determine the flux in the lab frame for
$k = x_1 \gamma_L m_p$, equivalent to the center of mass frame in a
collider.  The photon flux is exponentially suppressed for $k>\gamma_L
\hbar c/R_A$, corresponding to a momentum fraction $x_1 > \hbar
c/m_pR_A$.  The maximum $\gamma N$ center of mass energy,
$\sqrt{S_{\gamma N}}$, is much lower than the hadronic $\sqrt{S}$.  Note that
$\sqrt{S_{\gamma N}} = W_{\gamma N}$, the typical notation for HERA.  For
consistency, we use $\sqrt{S}$ notation for all three processes.

The cross section for direct photon-nucleon heavy quark
photoproduction is obtained by convoluting Eq.~(\ref{gamApart}) with the
photon flux and the gluon distribution in the nucleus and integrating over
$k$ and $x_2$,
\begin{equation}
S^2\frac{d^2\sigma^{\rm dir}_{\gamma A 
\rightarrow Q \overline Q X}}{dT_1 dU_1 d^2b} =
2 \int dz \int_{k_{\rm min}}^\infty dk {d^3N_\gamma \over 
dkd^2b} \int_{x_{2_{\rm min}}}^1 
\frac{dx_2}{x_2} F_g^A(x_2,Q^2,\vec b,z)  s^2 
\frac{d^2 \sigma_{\gamma g}}{dt_1 du_1} \, \, ,
\label{main}
\end{equation}
where $d^3N_\gamma/dkd^2b$ is the differential photon flux from one
nucleus (our final results will be integrated over $b>2R_A$) and $z$ is the
longitudinal distance.  The factor of
two in Eq.~(\ref{main}) arises because both nuclei emit photons and
thus serve as targets.  The incoherence of heavy quark production
eliminates interference between the two production
sources\cite{usinterf}.  Four-momentum conservation gives $x_{2_{\rm
min}} = -U_1/(S + T_1)$ in terms of the nucleon-nucleon invariants.
The equivalent hadronic invariants can be defined for photon four
momentum $k$ as $S_{\gamma N} = (k + P_2)^2$, $T_{1,\gamma N} = (P_2 -
p_1)^2 - m_Q^2$, and $U_{1, \gamma N} = (k - p_1)^2 - m_Q^2$
\cite{SvN}.  The partonic and equivalent hadronic invariants for fixed
$k$ are related by $s = x_2S_{\gamma N}$, $t_1 = U_{1, \gamma N}$, and
$u_1 = x_2 T_{1, \gamma N}$.

We now turn to the resolved (hadronic) contribution to the
photoproduction cross section.  The hadronic reaction, $\gamma N
\rightarrow Q \overline Q X$, is unchanged, but now, prior to the
interaction with the nucleon, the photon splits into a color singlet
state with some number of $q \overline q$ pairs and gluons.  There are
a few photon parton distributions
available~\cite{GRVgam,DG1,LAC1,WHIT,SaS}.  None of them can be
definitively ruled out by the existing data on the photon structure
function \cite{PDFLIB,JADE}.  As expected, $F_q^\gamma(x,Q^2) = F_{\overline
q}^\gamma (x,Q^2)$ flavor by flavor because there are no ``valence'' quarks in
the photon.  The gluon distribution in the photon is less well known.
We use the GRV LO set \cite{GRVgam}.  Its gluon distribution is
similar to most of the other available sets \cite{DG1,WHIT,SaS}.  Only
the LAC1 set \cite{LAC1} has a higher low-$x$
gluon density, up to an order of magnitude larger than the others.

On the parton level, the resolved LO reactions are $g(xk) +
g(x_2 P_2) \rightarrow Q(p_1) + \overline Q(p_2)$ and $q (xk) +
\overline q(x_2 P_2) \rightarrow Q(p_1) + \overline Q(p_2)$ where $x$ is the
fraction of the photon momentum carried by the parton.  The LO diagrams for
resolved photoproduction, shown in Fig.~\ref{Feynphot}(b)-(d), are the same as
for hadroproduction except that one parton source is a photon
rather than a nucleon.
The LO partonic cross sections are 
\cite{CTEQRMP}
\begin{eqnarray}
\hat{s}^2 \frac{d^2 \sigma_{q \overline q}}{d\hat{t}_1 d\hat{u}_1} & = 
& \pi \alpha_s^2(Q^2)
\frac{4}{9} \left( \frac{\hat{t}_1^2 + \hat{u}_1^2}{\hat{s}^2} 
+ \frac{2m_Q^2}{\hat{s}} \right) 
\delta(\hat{s} + \hat{t}_1 + \hat{u}_1) \, \, , \label{qqpartres} \\
\hat{s}^2 \frac{d^2 \sigma_{gg}}{d\hat{t}_1 d\hat{u}_1} & = & \frac{\pi 
\alpha_s^2(Q^2)}{16}
B_{\rm QED} (\hat{s},\hat{t}_1,\hat{u}_1) \left[ 3 \left( 1 - 
\frac{2\hat{t}_1 \hat{u}_1}{\hat{s}^2} \right)
- \frac{1}{3} \right] \delta(\hat{s} + \hat{t}_1 + \hat{u}_1)\, \, ,
\label{ggpartres}
\end{eqnarray}
where $\hat{s} = (xk + x_2P_2)^2$, $\hat{t}_1 = (xk - p_1)^2 - m_Q^2$,
and $\hat{u}_1 = (x_2P_2 - p_1)^2 - m_Q^2$.  The $gg$ partonic cross
section, Eq.~(\ref{ggpartres}), is proportional to the photon-gluon 
fusion cross section, Eq.~(\ref{gamApart}), with an additional factor 
for the non-Abelian
three-gluon vertex.  The $q \overline q$ annihilation cross section
has a different structure because it is an $s$-channel process with
gluon exchange between the $q \overline q$ and $Q \overline Q$
vertices.  The $gg$ reactions are shown in
Figs.~\ref{Feynphot}(b)-(c); Fig.~\ref{Feynphot}(c) is the non-Abelian
contribution.  The $q \overline q$ diagram is shown in
Fig.~\ref{Feynphot}(d). Modulo the additional factor in the $gg$ cross
section, the resolved partonic photoproduction cross sections are a
factor $\alpha_s(Q^2)/\alpha e_Q^2$ larger than the direct, $\gamma
g$, partonic photoproduction cross sections.  Despite this, the resolved 
component is still smaller than the direct component.

The cross section for resolved photoproduction is
\begin{eqnarray}
\lefteqn{S^2\frac{d^2\sigma^{\rm res}_{\gamma A \rightarrow Q \overline Q
X}}{dT_1 dU_1 d^2b} = 2 \int dz \int_{k_{\rm min}}^\infty 
\frac{dk}{k} {d^3N_\gamma\over dkdb^2} \int_{k_{\rm min}/k}^1 \frac{dx}{x}
\int_{x_{2_{\rm min}}}^1 \frac{dx_2}{x_2}} \nonumber \\
&  & \mbox{} \times \left[ F_g^\gamma (x,Q^2) 
F_g^A(x_2,Q^2,\vec b,z)  \hat{s}^2 
\frac{d^2 \sigma_{gg}}{d\hat{t}_1 d\hat{u}_1} \right. \nonumber  \\
&  & \mbox{} + \left. \sum_{q=u,d,s} F_q^\gamma (x,Q^2) 
\left\{
F_q^A(x_2,Q^2,\vec b,z) + F_{\overline q}^A(x_2,Q^2,\vec b,z) \right\}
\hat{s}^2 \frac{d^2 \sigma_{q \overline q}}{d\hat{t}_1 
d\hat{u}_1} \right] \, \, ,
\label{mainres}
\end{eqnarray}
where $k_{\rm min}$ is defined as before.  Since $k$ is typically
larger in resolved than direct photoproduction, the average photon
flux is lower in the resolved contribution.

The nuclear parton densities $F_i^A(x,Q^2,\vec{b},z)$ in Eqs.~(\ref{main}) and
(\ref{mainres}) can be
factorized into $x$ and $Q^2$ independent nuclear density
distributions, position and nuclear-number independent nucleon parton
densities, and a shadowing function $S^i(A,x,Q^2,\vec{b},z)$ that
describes the modification of the nuclear parton distributions in
position and momentum space.
Then \cite{spenprc,spenprl,spenpsi,ekkv4,RVwz}
\begin{eqnarray}
F_i^A(x,Q^2,\vec{b},z) & = & \rho_A(\vec{b},z) S^i(A,x,Q^2,\vec{b},z)
f_i^N(x,Q^2) \label{nucglu} \, \, 
\end{eqnarray}
where $f^N_i(x,Q^2)$ is the parton density in the nucleon.  We evaluate 
the MRST LO parton distributions \cite{pdf} at $Q^2= a^2 m^2_T$
where $a = 2$ for charm and 1 for bottom.  In the absence of nuclear
modifications, $S^i(A,x,Q^2,\vec{b},z)\equiv1$.  The nuclear density
distribution, $\rho_A(\vec{b},z)$, is a Woods-Saxon shape
with parameters determined from electron
scattering data \cite{Vvv}.  Although most models of shadowing predict
a dependence on the parton position in the nucleus, in this
photoproduction calculation we neglect any impact parameter
dependence.  Then the position dependence drops out of $S^i$.
We employ the
EKS98 shadowing parameterization \cite{eskola}, available in PDFLIB
\cite{PDFLIB} for $S^i \neq 1$.

The full photoproduction cross section is the sum of the direct and resolved
contributions \cite{frixione},
\begin{eqnarray}
S^2\frac{d^2\sigma_{\gamma A \rightarrow Q \overline Q 
X}}{dT_1 
dU_1 d^2b} = S^2\frac{d^2\sigma^{\rm dir}_{\gamma A \rightarrow Q 
\overline Q X}}{dT_1 dU_1 d^2b} + S^2\frac{d^2\sigma^{\rm res}_{\gamma A 
\rightarrow Q \overline Q X}}{dT_1 dU_1 d^2b} \, \, .
\label{phottot}
\end{eqnarray}
The total cross section is the integral over impact parameter and the hadronic
invariants $T_1$ and $U_1$, 
\begin{eqnarray}
\sigma_{\gamma A \rightarrow Q \overline Q X} = \int dT_1 \, dU_1
\, d^2b \, \frac{d^2\sigma_{\gamma A \rightarrow Q \overline Q 
X}}{dT_1 dU_1 d^2b} \, \, .
\label{photinttot}
\end{eqnarray}
When $S^i=1$, the impact-parameter integrated cross section in
Eq.~(\ref{photinttot}) scales with $A$.  Including shadowing makes the
dependence on $A$ nonlinear.

The photon flux is given by the Weizs\"acker-Williams method.  The
flux from a charge $Z$ nucleus a distance $r$ away is
\begin{equation}
{d^3N_\gamma \over dkd^2r} = 
{Z^2\alpha w^2\over \pi^2kr^2} \left[ K_1^2(w) + {1\over
\gamma_L^2} K_0^2(w) \right] \, \, 
\label{wwr}
\end{equation}
where $w=kr/\gamma_L$ and $K_0(w)$ and $K_1(w)$ are modified Bessel
functions.  The photon flux decreases exponentially above a cutoff
energy determined by the size of the nucleus.  In the lab frame, the
cutoff is $k_{\rm max} \approx \gamma_L \hbar c/R_A$.  In the rest frame of the
target nucleus, the cutoff is boosted to 
$E_{\rm max}=(2\gamma_L^2-1)\hbar c/R_A$.  Table~\ref{gamfacs} shows the
beam energies, $E_{\rm beam}$, Lorentz factors, $\gamma_L$, $k_{\rm
max}$, and $E_{\rm max}$, as well as the corresponding maximum center
of mass energy, $\sqrt{S_{\rm max}} = \sqrt{2E_{\rm max}m_p}$, for single
photon interactions with protons, $\gamma p \rightarrow Q \overline Q$
\cite{frstr}.  At the LHC, the energies are high enough for $t
\overline t$ photoproduction\cite{knv1}.

The total photon flux striking the target nucleus is the integral of
Eq.~(\ref{wwr}) over the transverse area of the target for all impact
parameters subject to the constraint that the two nuclei do not
interact hadronically\cite{usPRC}.  This must be calculated
numerically.  However, a reasonable analytic approximation for $AA$
collisions is given by the photon flux integrated over radii $r>2R_A$.
The analytic photon flux is
\begin{equation}
{dN_\gamma\over dk} = 
{2Z^2 \alpha \over\pi k} \left[ w_R^{AA}K_0(w_R^{AA})
K_1(w_R^{AA})- {(w_R^{AA})^2\over 2} \big(K_1^2(w_R^{AA})-K_0^2(w_R^{AA})
\big) \right] \, \, 
\label{analflux}
\end{equation}
where $w_R^{AA}=2kR_A/\gamma_L$.  We use the more accurate numerical
calculations here. The difference between the numerical and analytic
expressions is typically less than 15\%, except for photon energies
near the cutoff.  The analytical and numerical photon fluxes differ most
for $b\overline b$ production at RHIC.  

\begin{figure}
\setlength{\epsfxsize=0.95\textwidth}
\setlength{\epsfysize=0.5\textheight}
\centerline{\epsffile{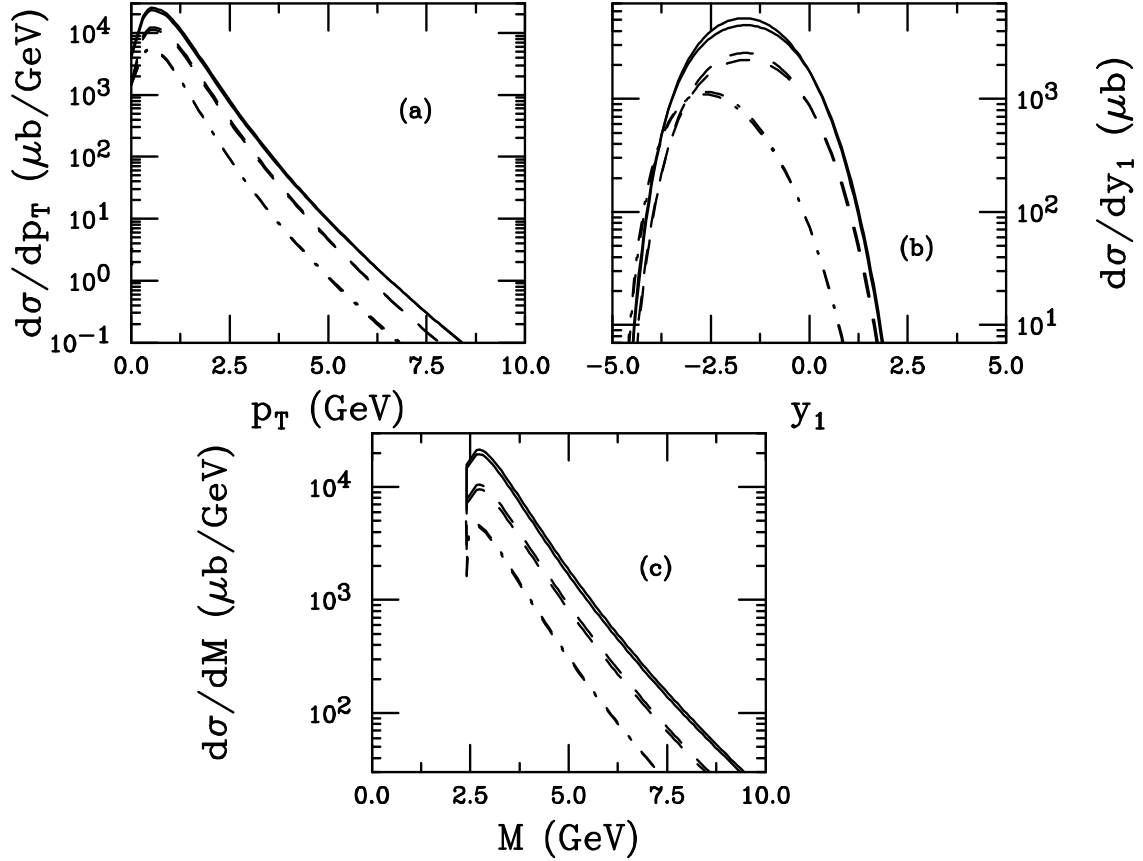}}
\caption[]{Charm photoproduction in peripheral Au+Au collisions at
RHIC for $b> 2R_A$.  The single $c$ quark
$p_T$ (a) and rapidity (b) distributions are shown along with the $c
\overline c$ pair invariant mass (c).  The direct (dashed), resolved
(dot-dashed), and the sum of the two (solid) are shown.  The direct
contribution is divided by two to distinguish it from the total while
the resolved contribution is multiplied by ten.  There are two curves
for each contribution: $S^i =1$ and EKS98.  At this energy, the curves
are almost indistinguishable but the curves with shadowing are somewhat
higher, especially at negative rapidities.  In the rapidity distributions, the
photon is coming from the left.}
\label{gamArc}
\end{figure}

\begin{figure}
\setlength{\epsfxsize=0.95\textwidth}
\setlength{\epsfysize=0.5\textheight}
\centerline{\epsffile{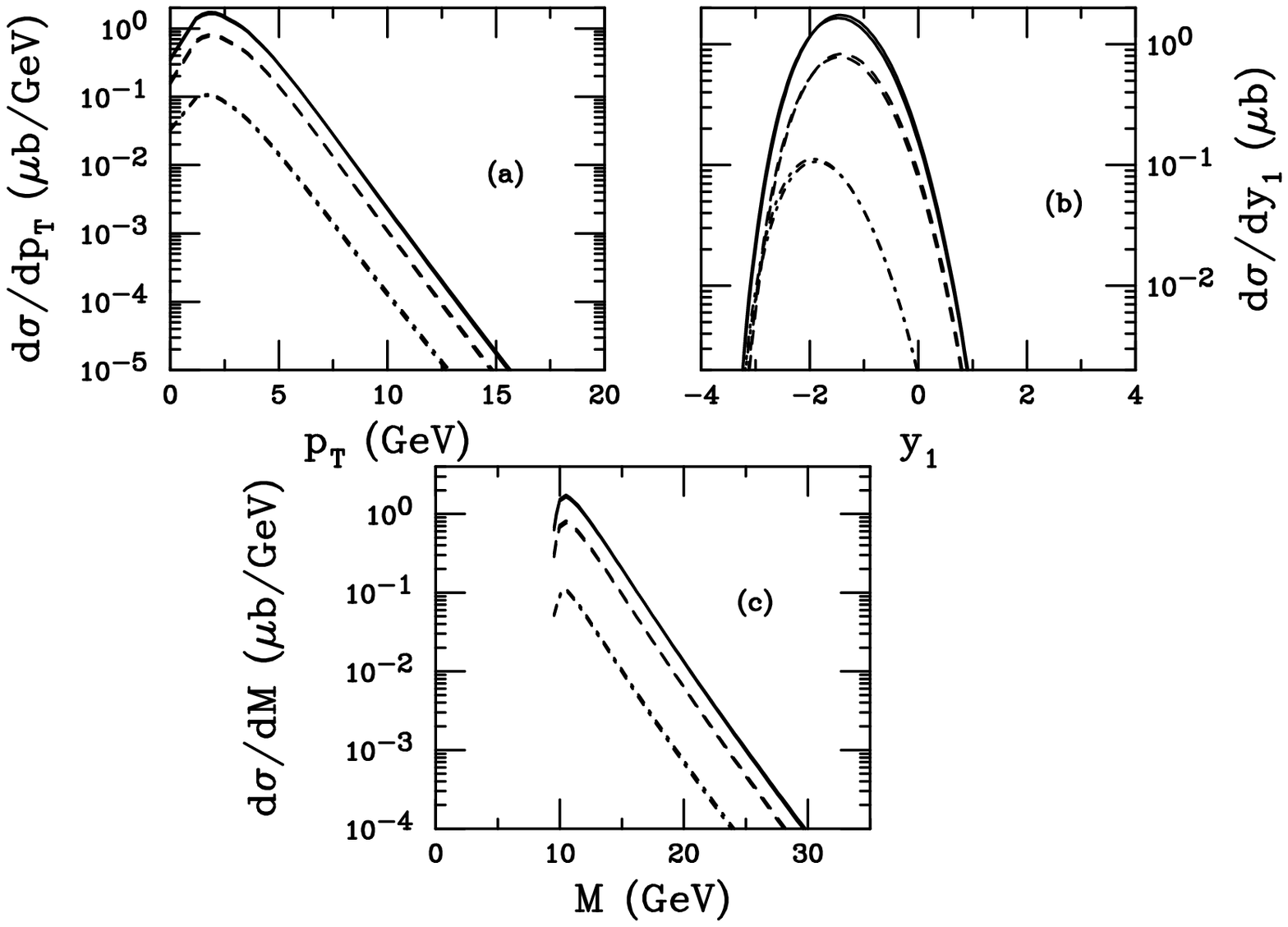}}
\caption[]{Bottom photoproduction in peripheral Au+Au collisions at
RHIC for
$b>2R_A$.  The single $b$ quark $p_T$ (a) and rapidity (b)
distributions are shown along with the $b \overline b$ pair invariant
mass (c).  The direct (dashed), resolved (dot-dashed), and the sum of
the two (solid) are shown.  The direct contribution is divided by two
to distinguish it from the total.  There are two curves for each
contribution: $S^i =1$ and EKS98.  At this energy, the curves are
almost indistinguishable.  In the rapidity distributions, the photon
is coming from the left.}
\label{gamArb}
\end{figure}

\begin{figure}
\setlength{\epsfxsize=0.95\textwidth}
\setlength{\epsfysize=0.5\textheight}
\centerline{\epsffile{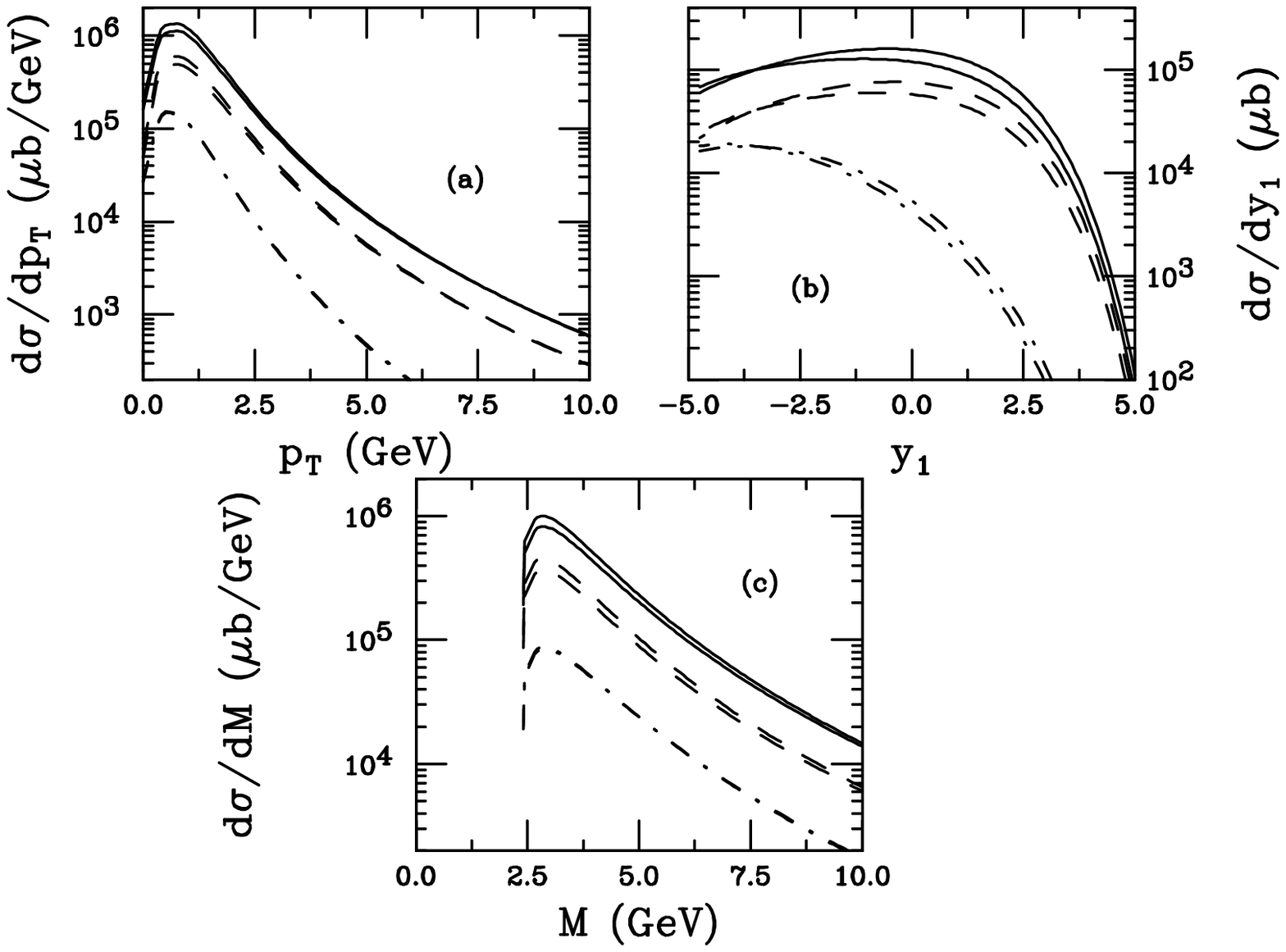}}
\caption[]{Charm photoproduction in peripheral Pb+Pb collisions at LHC
for $b> 2R_A$.  The single $c$ quark
$p_T$ (a) and rapidity (b) distributions are shown along with the $c
\overline c$ pair invariant mass (c).  The direct (dashed), resolved
(dot-dashed), and the sum of the two (solid) are shown.  The direct
contribution is divided by two to distinguish it from the total.
There are two curves for each contribution: $S^i =1$ and EKS98.  The
unshadowed curves are higher than the shadowed, particularly at large
rapidities.  In the rapidity
distributions, the photon is coming from the left.}
\label{gamAlc}
\end{figure}

\begin{figure}
\setlength{\epsfxsize=0.95\textwidth}
\setlength{\epsfysize=0.5\textheight}
\centerline{\epsffile{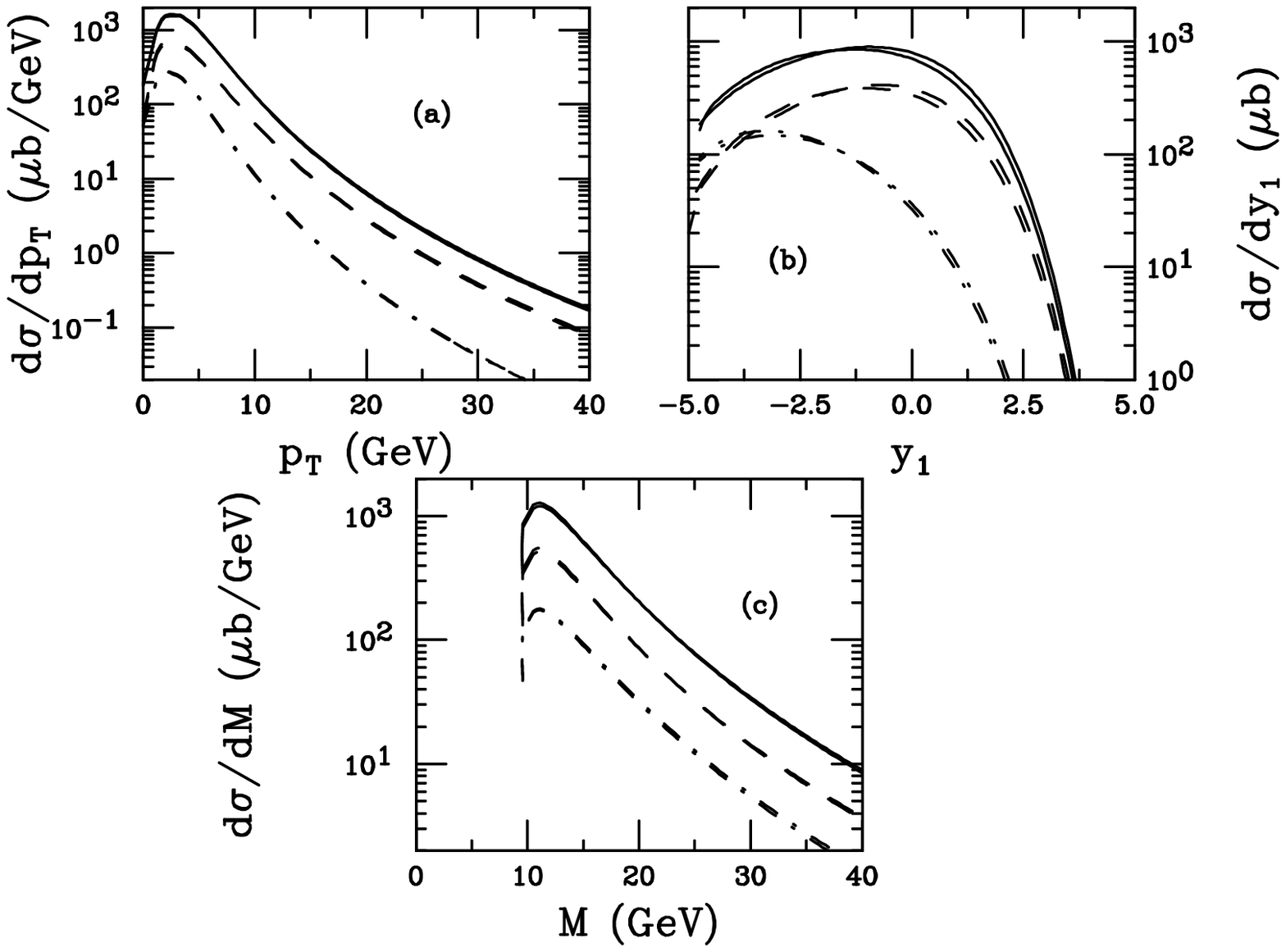}}
\caption[]{Bottom photoproduction in peripheral Pb+Pb collisions at
LHC for $b> 2R_A$.  The single $b$ quark
$p_T$ (a) and rapidity (b) distributions are shown along with the $b
\overline b$ pair invariant mass (c).  The direct (dashed), resolved
(dot-dashed), and the sum of the two (solid) are shown.  The direct
contribution is divided by two to distinguish it from the total.
There are two curves for each contribution: $S^i =1$ and EKS98.  The
unshadowed curves are higher than the shadowed.  In the rapidity
distributions, the photon is coming from the left.}
\label{gamAlb}
\end{figure}

The photoproduction distributions are shown in
Figs.~\ref{gamArc}-\ref{gamAlb} for the largest
nuclei at each energy, gold for RHIC and lead for LHC.  Since
$\sqrt{S_{\rm max}}$ is close to the $b \overline b$ production
threshold for iodine and gold beams at RHIC, the $p_T$ and mass
distributions for these nuclei are narrower than those with oxygen and
silicon beams.  The direct photoproduction
results are reduced by a factor of two on the figures to separate them
from the total.  There are two curves for each contribution, one
without shadowing, $S^i = 1$, and one with homogeneous nuclear
shadowing, $S^i =$ EKS98.  When the effects of shadowing are small,
the curves are indistinguishable.

Shadowing has the largest effect on the rapidity distributions in
Figs.~\ref{gamArc}(b)-\ref{gamAlb}(b).  In these calculations, the
photon is emitted from the nucleus coming from positive rapidity.  Then
$y_1<0$ corresponds to $k<\gamma_L x_2m_p$ in the center of mass (lab)
frame.  If the photon emitter and target nucleus are interchanged, the
resulting unshadowed rapidity distribution, $S^i=1$, is the mirror
image of these distributions around $y_1=0$. The
$Q$ and $\overline Q$ distributions are asymmetric around $y_1=0$.
The resolved contribution is largest at rapidities where the photon
momentum is small. The resolved rapidity distributions are narrower
and shifted to larger negative rapidity than the direct contribution.
The average mass and transverse momentum for the resolved component
are smaller than for the direct ones.  The total heavy quark rapidity
distributions are the sum of the displayed results with their mirror
images when both nuclei emit photons.  This factor of two is included
in the transverse momentum and invariant mass distributions.

The impact-parameter integrated total direct and resolved
photoproduction cross sections are given in Table~\ref{gamAcc} for
charm and Table~\ref{gamAbb} for bottom.  The difference between the
$S^i = 1$ and $S^i$ = EKS98 calculations is due to
shadowing.  The change is about 10\% for $c\overline c$ with gold at RHIC,
rising to 20\% for $c\overline c$ with lead at the LHC.  For
$b\overline b$, the shadowing effect is smaller, about 5\%.  By selecting charm
production in a limited rapidity and $p_T$ range, it would be possible
to enhance the effect of shadowing slightly but the 10\% and 20\% effects at
RHIC and the LHC are useful benchmarks of the accuracy needed for 
a meaningful measurement.

The cross sections vary by orders of magnitude between the lightest
and heaviest targets, primarily due to the changing $Z^2$.  
At RHIC, $(Z_{\rm Au}/Z_{\rm O})^2$ is almost exactly the ratio of the
$b \overline b$ production cross sections.  For $b\overline b$ production 
at the LHC, the lead to oxygen cross section ratio is $\sim 60$, less 
than $(Z_{\rm Pb}/Z_{\rm O})^2$.

After adjustment for different parton distributions, quark masses, beam
energies and species as well as the resolved contributions, the
photoproduction results are comparable with previous studies.  The
direct $c \overline c$ cross sections in Table~\ref{gamAcc} are
almost identical to those found by Baron and Baur\cite{bbar}, despite
significant differences in gluon structure function and quark masses
({\it e.g.}\ $m_c=1.74$ GeV in Ref.~\cite{bbar} rather than $m_c=1.2$
GeV).  The quark mass difference can be compensated by the newer parton
distributions since the low-$x$ MRST gluon density is much higher than 
that of the older Duke-Owens parton distributions used by Baron and Baur.
Our $b \overline b$ cross section is about three times higher than
Greiner {\it et al.} \cite{bbar2,hofmann} at the same invariant mass
although they use a smaller $b$ quark mass, 4.5 GeV,
due to the larger low-$x$ MRST gluon distribution.

On the other hand, nonperturbative QCD calculations of 
direct photoproduction can yield
very different rates. A colored glass model predicts $Q\overline Q$
production cross sections in Pb+Pb collisions at the LHC of 800 mb for
$c \overline c$ and 100 mb for $b \overline b$ \cite{glass}, about half the
perturbative $c \overline c$ cross section but 140 times the $b \overline b$
cross section.  The ratio of charm to bottom cross sections in this
formulation is extraordinarily small compared to the perturbative
results in Tables~\ref{gamAcc} and \ref{gamAbb}.

We now discuss the resolved rates.  At RHIC the $c \overline c$ and $b
\overline b$ resolved contributions are $\sim 2$ and 6\% of the total.
The contribution is small because the available $\gamma p$ energy is
relatively close to threshold:  the resolved center of
mass energy is smaller than $\sqrt{S_{\rm max}}$.  
At the LHC, the resolved contributions are
$\sim 15$ and 20\% of the total charm and bottom photoproduction cross
sections respectively, comparable to the shadowing effect.
Interestingly, the resolved $q \overline q$ component is
considerably greater than $q \overline q$ annihilation in hadroproduction 
because, like the valence
quark distributions of the proton, the photon quark and antiquark
distributions peak at large $x$.  However, the peak of the photon quark
distribution is at higher $x$ than in the proton.  These large photon 
momentum fractions increase the $q \overline q$ contribution near threshold.  
The total $c \overline c$ resolved cross
section at RHIC is $(35-50)$\% $q \overline q$ while the $b \overline
b$ resolved contribution is $(80-90)$\% $q \overline q$.  The relative
resolved contribution remains larger than $q \overline q$ annihilation
in hadroproduction at the LHC where the $q \overline q$ contribution
to the hadroproduction cross section is 4\% for charm and 10\% for
bottom.  
The LAC1 gluon distribution\cite{LAC1} predicts a considerably
larger resolved contribution with a smaller relative
$q\overline q$ contribution.

At fixed target energies, near threshold, the resolved component of
charm photoproduction is relatively unimportant\cite{E691}.  However,
at higher $\gamma p$ energies such as those at HERA, the resolved
component becomes important.  RHIC is an intermediate case, with an
average $\gamma p$ energy less than 1/5 of that at HERA.  We find
that, at RHIC, the resolved component is a $(2-6)$\% effect.  At
the LHC, where the energies are generally higher than at HERA, it is
$(15-20)$\% of the total cross section, concentrated at
large rapidities.  As was previously mentioned, much early
photoproduction data favors heavier quark masses, $1.5-1.8$ GeV,
although newer HERA results are ambiguous\cite{frix2}.  Because we use
a lower $m_c$, our cross sections are higher than a direct comparison
to HERA would indicate.  Changing $m_c$ affects the overall cross
section and the quark $p_T$ spectrum for $p_T < 5$ GeV\cite{frixione}.

As is typically done, we include all $Q \overline Q$ pairs in the
total cross sections and rates even though some of these pairs have
masses below the $H \overline H$ threshold where $H \overline H \equiv
D \overline D$, $B \overline B$.  Photoproduction is an inclusive
process; accompanying particles can combine with the $Q$ and
$\overline Q$, allowing the pairs with $M<2m_H$ to hadronize.  We
assume the hadronization process does not affect the rate.  Including
all pairs in the total cross section is presumably an even safer
assumption for hadroproduction because there are more accompanying
particles.  On the other hand, $\gamma \gamma$ interactions should
have no additional particles since the interaction is purely
electromagnetic and occurs away from the nuclei in free space.
Section \ref{gamgamsec} will discuss this in more detail.

One way to avoid some uncertainties due to higher order corrections is
to measure shadowing by comparing the $pA$ and $AA$ photoproduction
cross sections at equal photon energies since the parameter dependence
cancels in the ratio $\sigma(AA)/\sigma(pA)$.  In the equal speed
system, equal photon energies correspond to the same final-state
rapidities.  In $pA$ collisions, the photon almost always comes from
the nucleus due to its stronger field.  Thus the $pA$ rates 
depend on the free proton gluon
distribution.  The photon fluxes are different for $pA$ and $AA$
because the minimum radii used to determine $\omega_R$ are different:
$2R_A$ in $AA$ compared to $R_A+r_p$ in $pA$ where $r_p$ is the proton 
radius.  There are a number of ways to define the proton radius.  We use
the hadronic radius, $r_p\approx 0.6$ fm, determined from
photoproduction data\cite{herarho}.  As we will show, our results are
not very sensitive to $r_p$.

In a detailed calculation, the hadronic interaction
probability near the minimum radius depends on the matter distribution
in the nucleus.  Our calculations use Woods-Saxon distributions with
parameters fit to electron scattering data.  This data is quite
accurate.  However, electron scattering is only sensitive to the
charge distribution in the nucleus.  Recent measurements indicate that
the neutron and proton distributions differ in nuclei \cite{pbars}.
This uncertainty in the matter distribution is likely to limit the
photon flux determination.

The uncertainty in the photon flux can be reduced by
calibrating it with other measurements.  Further studies of the matter
distributions in nuclei \cite{pbars} and the total ion-ion cross
sections, as well as comparisons with known photoproduction and
two-photon processes, can help pin down the photon flux.  For example,
the cross section for $\gamma p\rightarrow\rho p$ is known to 10\%
from measurements at HERA\cite{herarho}.  The $\gamma A\rightarrow VA$
data are limited to energies lower than but still
comparable to those available at RHIC.  A combination of lower energy
RHIC data and a judicious extrapolation could fix the calibration.
Studies of well known two-photon processes, like lepton production,
can also help refine the determination of the photon flux.  With these
checks, it should be possible to understand the photon flux in $pA$
relative to $AA$ to better than 10\%, good enough for a useful shadowing
measurement.

Our $pA$ results are calculated in the equal-speed frame.  This is
appropriate for RHIC where $\sqrt{S}$ is the same in $pA$ and $AA$
interactions.  At the LHC, the proton and nuclear beams must have the
same magnetic rigidity and, hence, different velocities and
per-nucleon energies.  At the LHC, the equal-speed frame is moving in the
laboratory frame so that the maximum $pA$ energy is larger than the $AA$
energy.  The $\gamma_L$ and $\sqrt{S}$ given for $pA$ at the LHC in
Tables~\ref{lum} and \ref{gamfacs} are those of the equal-speed
system.  The $pA$ total cross sections for $Q \overline Q$ production 
are given in Table~\ref{gampccbb}.

It is easy to compare $pA$ and $AA$ results at RHIC.  For $S^i=1$, the
only difference is in the impact parameter range: $b>2R_A$ in $AA$ compared to
$b>R_A+r_p$ in $pA$.  We compare the numerical $AA$ results presented in
Tables~\ref{gamAcc} and \ref{gamAbb} with those using the analytical
photon flux for $pA$ in Table~\ref{gampccbb}.  To normalize the photon
cross sections, we divide the $AA$ cross section by $2A$ because there
is only a single photon source in $pA$ and the proton target is a single
nucleon.  Due to the reduced minimum impact parameters, $\sigma(AA)/2A$ is
$(22-37)$\% lower for $c \overline c$ and $(37-65)$\% lower for $b \overline 
b$ with $r_p = 0.6$ fm.  The results differ least for small $A$
where $R_A - r_p$ is reduced.  Increasing $r_p$ by 50\% would
decrease the $c \overline c$ cross sections $(6-3.5)$\% and the $b \overline
b$ cross sections by $(11-7)$\%.  Changing $r_p$ has the largest effect for
small $A$ since $R_A/r_p$ is smaller.  The $b
\overline b$ differences are largest because the energy is close to
production threshold and high $k$ corresponds to small impact parameter.

At the LHC, the maximum $pA$ and $AA$ energies differ, adding another
variable to the comparison.  Here, we compare $\sigma(AA)/2A$ directly
with $\sigma(pA)$, using the analytic flux for both to maximize the
parallelism.  For equal $pA$ and $AA$ energies, the different $b$ range
is less significant at the LHC.  The decrease in $AA$
relative to $pA$ is $(18-23)$\% for $c \overline c$ and $(24-30)$\%
for $b \overline b$.  As at RHIC, the effects are larger for heavier
nuclei.  Increasing $r_p$ by 50\% is only a $(2-5)$\% effect with the 
bigger change for $b \overline b$.

The effect of the higher $pA$ energy is significant.  At the 
lower $AA$ beam energy,
the $pA$ cross section is reduced $(23-32)$\% for $c\overline c$ and
$(30-40)$\% for $b\overline b$ relative to the maximum $pA$
energy.  The major difference between the $c \overline c$ and $b
\overline b$ results is likely due to the smaller $x$ values probed in
$c \overline c$ production.  The increased energy has a larger effect
on the cross section than the change in the minimum impact
parameter. The energy dependence illustrates the desirability of
obtaining energy excitation functions in both $pA$ and $AA$ interactions.

Tables~\ref{NgamAccbb} and~\ref{Ngampccbb} give the total monthly $AA$
and $pA$ $Q \overline Q$ production rates assuming a $10^6$ s/month run. 
The $c \overline c$ rates
are quite high.  Of course, hadronization, branching ratios and
detector acceptances will all reduce the observed rates.  The $b
\overline b$ rates are only significant at LHC.  However, RHIC is a
dedicated heavy ion facility, originally
planned for more than 30 weeks of operation per
year compared to 4 weeks of heavy ion running at the LHC and should
thus accumulate more data than the tables indicate.

\section{Hadroproduction}
\label{hadrosec}

Hadroproduction of heavy quarks in heavy ion collisions has been
considered by many authors, see {\it e.g.}\,
\cite{spenprc,spenprl,vjmr,gmrv} and references therein.  Here, we
consider the special characteristics of heavy quark production in
peripheral heavy ion collisions.  Two aspects meriting our attention
are the form of the parton densities at the nuclear surface and the
overall appearance of the collision --- could hadroproduction 
mimic a photon-nucleus or two-photon interaction?  This
section addresses the first issue while section \ref{exsec} 
considers the second.

At leading order, heavy quarks are produced via the reaction $N(P_1) +
N(P_2) \rightarrow Q(p_1) + \overline Q(p_2) + X$.  The LO parton
reactions are $g(x_1 P_1) + g(x_2 P_2) \rightarrow Q(p_1) + \overline
Q(p_2)$ and $q (x_1 P_1) + \overline q(x_2 P_2) \rightarrow Q(p_1) +
\overline Q(p_2)$.  The LO partonic cross sections are those of
Eqs.~(\ref{qqpartres}) and (\ref{ggpartres}) with $\hat{s}$,
$\hat{t}_1$, and $\hat{u}_1$ replaced by $s$, $t_1$ and $u_1$.  Thus the
partonic couplings are a factor $\alpha_s(Q^2)/\alpha e_Q^2$ larger than 
direct photoproduction but have the same coupling strength as resolved 
photoproduction, as discussed in the previous section.  The partonic 
invariants, $s$, $t_1$, and $u_1$, are now $s = (x_1P_1 + x_2 P_2)^2$, $t_1 =
(x_1P_1 - p_1)^2 - m_Q^2 = (x_2 P_2 - p_2)^2 - m_Q^2$, and $u_1 = (x_2
P_2 - p_1)^2 - m_Q^2 = (x_1P_1 - p_2)^2 - m_Q^2$.

The hadronic heavy quark production cross
section is the convolution of two nuclear parton distributions with the 
partonic cross sections,
\begin{eqnarray}
S^2\frac{d^2\sigma_{A A \rightarrow Q \overline Q X}}{dT_1 dU_1 d^2b d^2r} 
& = & \int dz \, dz' \frac{dx_1}{x_1} \frac{dx_2}{x_2} \left(
F_g^A(x_1,Q^2,\vec r,z) F_g^B(x_2,Q^2,|\vec b - \vec r|,z') s^2 \frac{d^2 
\sigma_{gg}}{dt_1 du_1}  \right. \nonumber \\ 
&  & \left. \mbox{} + \sum_{q=u,d,s}  
\left\{ F_q^A(x_1,Q^2,\vec r, z) F_{\overline 
q}^B(x_2,Q^2,|\vec b - \vec r|,z') \right. \right. \nonumber \\ 
&  & \left. \left. \mbox{} + F_{\overline q}^A(x_1,Q^2,\vec r, z) 
F_q^B(x_2,Q^2,|\vec b - \vec r|,z') 
\right\} s^2 \frac{d^2 \sigma_{q \overline q}}{dt_1 du_1} \right) \, \, .
\label{pphad}
\end{eqnarray}
If the quark is detected, the hadronic invariants are again $S = (P_1
+ P_2)^2$, $T_1 = (P_2 - p_1)^2 - m_Q^2$, and $U_1 = (P_1 - p_1)^2 -
m_Q^2$ \cite{CTEQRMP}.  The partonic and hadronic invariants are now
related by $s = x_1x_2S$, $t_1 = x_1U_1$, and $u_1 = x_2 T_1$.
Four-momentum conservation at leading order gives $x_{2_{\rm min}} =
-x_1U_1/(x_1S + T_1)$.  We again perform a fully leading order
calculation using the MRST LO parton distributions and a one-loop
evaluation of $\alpha_s(Q^2)$.

Since both the projectile and target partons come from nuclei, we
consider the spatial dependence of the nuclear parton distributions, as 
suggested by shadowing measurements \cite{e745}, in
more detail.  Shadowing in peripheral collisions could then be significantly
different than in central collisions.  We consider three 
scenarios: no shadowing, $S^i = 1$; homogeneous shadowing with the
EKS98 parameterization, $S^i$ = EKS98; and inhomogeneous shadowing,
$S^i$ = EKS98$b$, with spatial dependence included, as
described in Refs.~\cite{spenprc,spenprl,spenpsi,ekkv4,RVwz}.

To obtain the $AA$ cross sections, we integrate Eq.~(\ref{pphad}) over
$d^2r$ and $d^2b$.  When $S^i = 1$ or EKS98, the only $b$ dependence
is in the nuclear density distributions and
\begin{eqnarray}
\sigma_{AA} \propto \sigma_{NN} \int d^2b \, d^2r \, T_A(r) \, T_A(|\vec b
- \vec r|) \, \, \label{tab}
\end{eqnarray}
where $T_A(r) = \int dz \rho_A(\vec r, z)$ is the nuclear profile function 
and $\sigma_{NN}$ is the $Q \overline Q$ production cross section in an
$NN$ collision.  When Eq.~(\ref{pphad}) is integrated over all $b$, 
$\sigma_{AA} \propto \sigma_{NN} A^2$ for
$S^i=1$.  Homogeneous shadowing makes the $A$ dependence nonlinear
so that  the integrated cross section is
effectively $\sigma_{AA} \propto \sigma_{NN} A^{2\alpha}$ where $\alpha$
can be determined from the $A$ dependence of the EKS98 parameterization.
Of course when $S^i$ = EKS98$b$, the full integral over $b$ and $r$ 
is needed in Eq.~(\ref{pphad}). In this section the
impact parameter integral is restricted to $b > 2R_A$ to be consistent
with the photoproduction results.

This calculation treats the nuclei as a continuous fluid, neglecting
the lumpiness due to the individual nucleons. When $b>2R_A$, only a
handful of nucleon-nucleon collisions can occur, resulting in
significant statistical fluctuations.  These fluctuations, although
unimportant for heavy quark production, may
affect some of the observables used for impact parameter determination
such as transverse energy or charged particle multiplicity.

\begin{figure}
\setlength{\epsfxsize=0.95\textwidth}
\setlength{\epsfysize=0.5\textheight}
\centerline{\epsffile{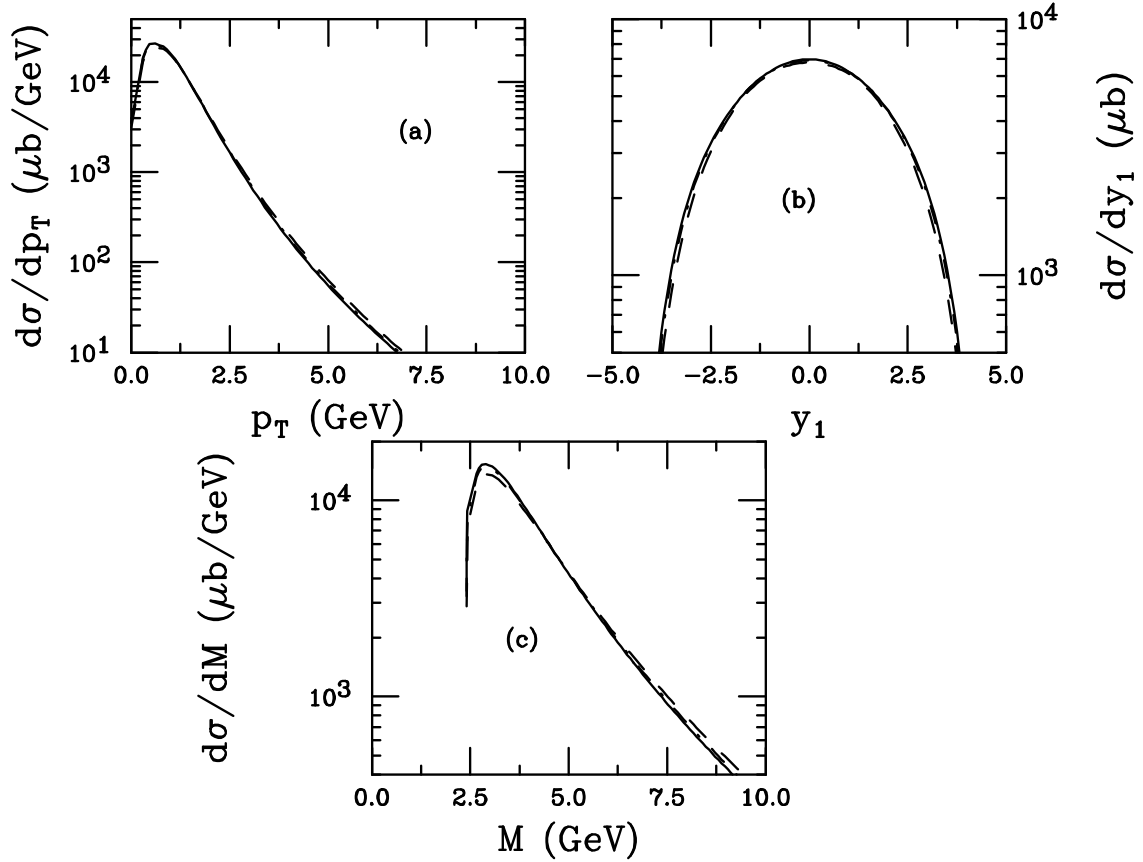}}
\caption[]{Charm hadroproduction in peripheral Au+Au collisions at RHIC
for $b> 2R_A$.  The single $c$ quark $p_T$ (a)
and rapidity (b) distributions are shown along with the $c \overline c$ pair
invariant mass (c).  The curves are $S^i = 1$ (solid), EKS98 (dashed), 
and EKS98$b$ (dot-dashed).  At this energy, the three curves are almost
indistinguishable.}
\label{qqAArc}
\end{figure}

\begin{figure}
\setlength{\epsfxsize=0.95\textwidth}
\setlength{\epsfysize=0.5\textheight}
\centerline{\epsffile{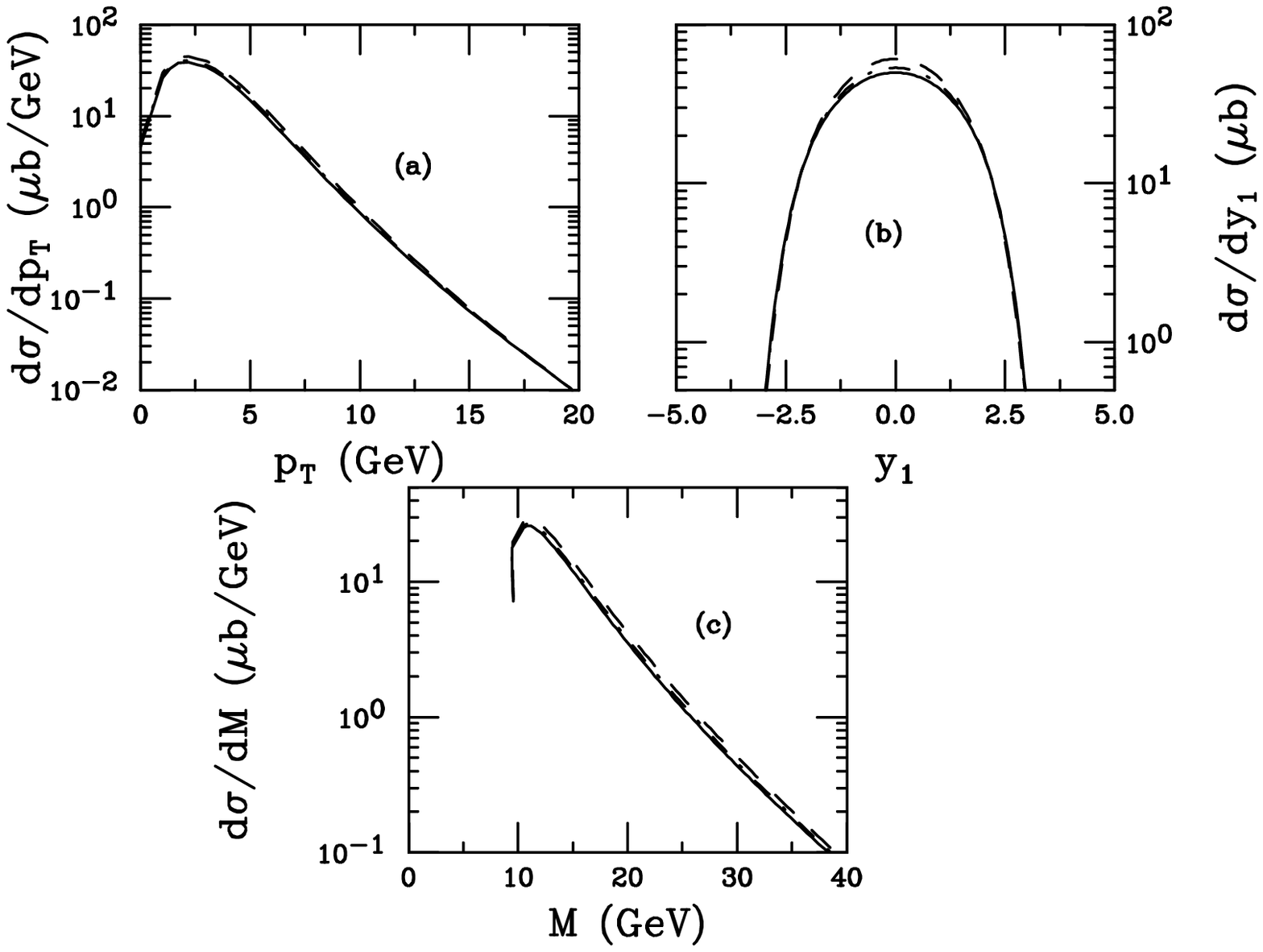}}
\caption[]{Bottom hadroproduction in peripheral Au+Au collisions at RHIC
for $b>2R_A$.  The single $b$ quark $p_T$ (a)
and rapidity (b) distributions are shown along with the $b \overline b$ pair
invariant mass (c).   The curves are $S^i = 1$ (solid), EKS98 (dashed), 
and EKS98$b$ (dot-dashed).}
\label{qqAArb}
\end{figure}

\begin{figure}
\setlength{\epsfxsize=0.95\textwidth}
\setlength{\epsfysize=0.5\textheight}
\centerline{\epsffile{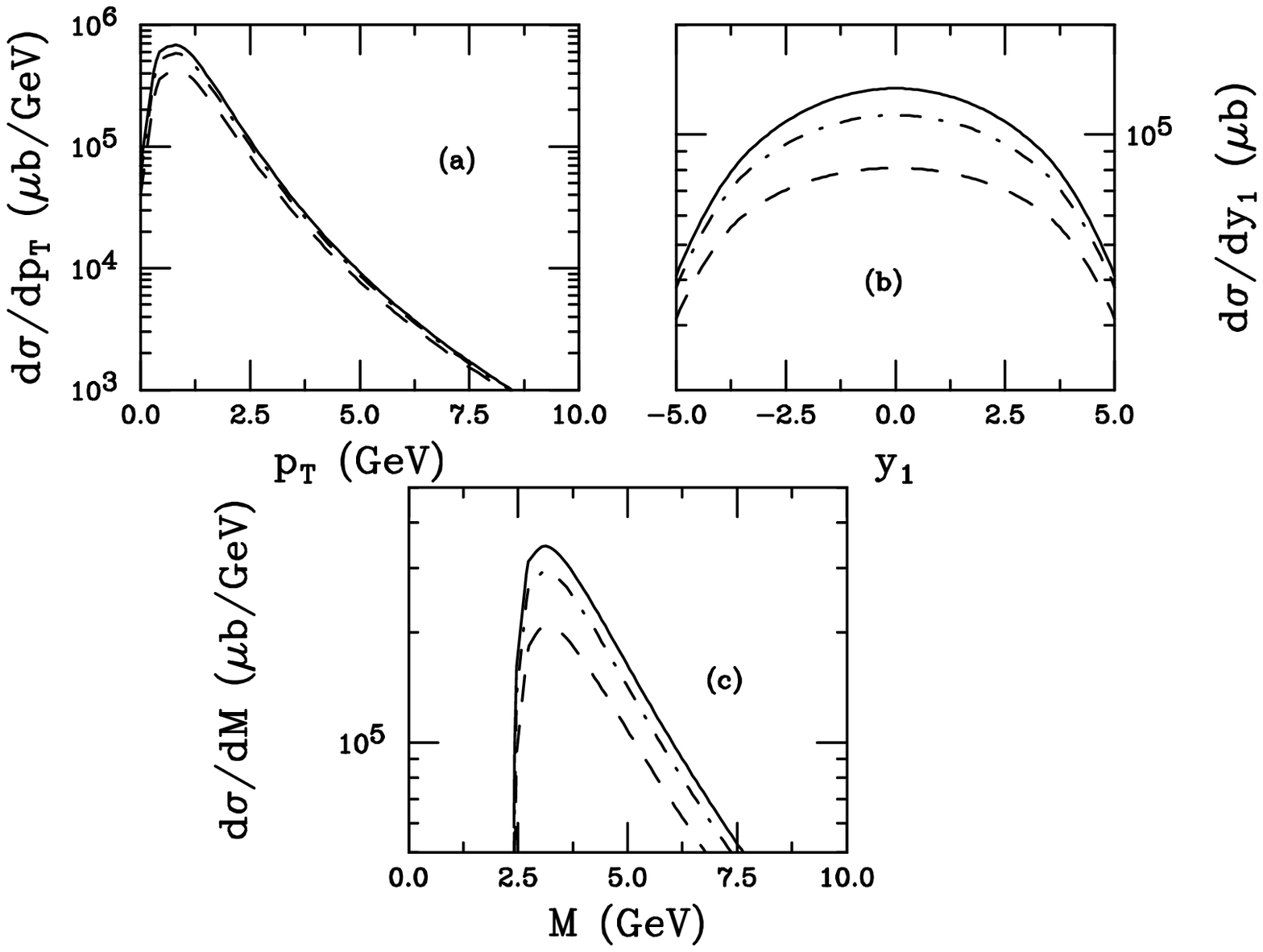}}
\caption[]{Charm hadroproduction in peripheral Pb+Pb collisions at LHC
for $b>2R_A$.  The single $c$ quark $p_T$ (a)
and rapidity (b) distributions are shown along with the $c \overline c$ pair
invariant mass (c).  The curves are $S^i = 1$ (solid), EKS98 (dashed), 
and EKS98$b$ (dot-dashed).}
\label{qqAAlc}
\end{figure}

\begin{figure}
\setlength{\epsfxsize=0.95\textwidth}
\setlength{\epsfysize=0.5\textheight}
\centerline{\epsffile{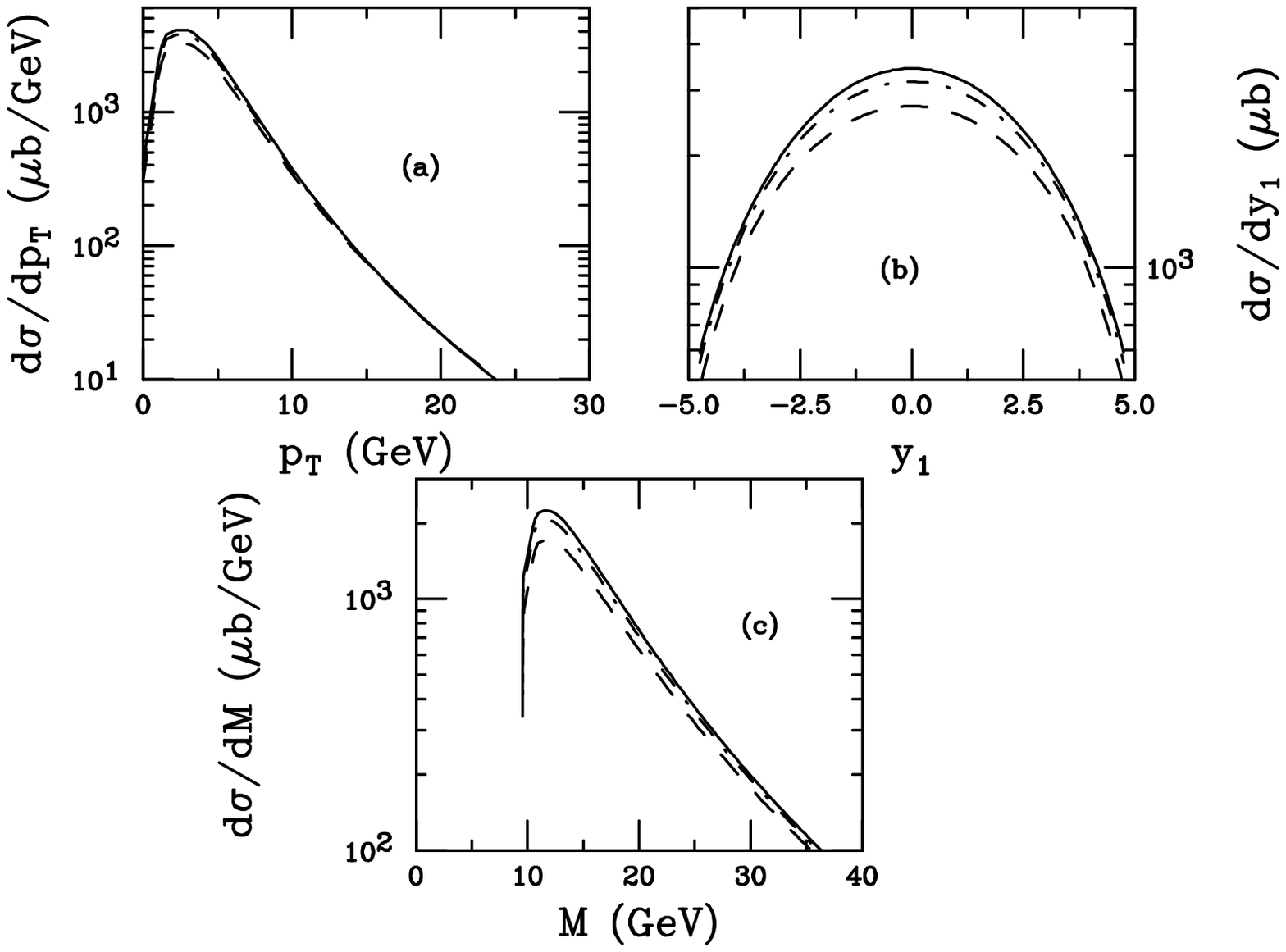}}
\caption[]{Bottom hadroproduction in peripheral Pb+Pb collisions at LHC
for $b>2R_A$.  The single $b$ quark $p_T$ (a)
and rapidity (b) distributions are shown along with the $b \overline b$ pair
invariant mass (c).  The curves are $S^i = 1$ (solid), EKS98 (dashed), 
and EKS98$b$ (dot-dashed).}
\label{qqAAlb}
\end{figure}

The hadroproduction distributions for the heaviest ions are shown in
Figs.~\ref{qqAArc}-\ref{qqAAlb}.  There are 
three curves for each colliding system.  When the small $x$ region is
probed, the $S^i=1$ result is the highest, EKS98 is the lowest, and 
EKS98$b$, with $b>2R_A$, 
is between the other two.  If we consider minimum impact parameters much
larger than $2R_A$, the EKS98$b$ curves would move even closer to 
the $S^i = 1$ results.  The order is reversed for $b \overline b$ production at
RHIC because $x \sim M/\sqrt{S} \sim 0.05-0.1$ for $2m_b < M < 20$ GeV, in
the gluon antishadowing region of the EKS98 parameterization.  Thus
the shadowed results lie above those with $S^i=1$.  The quark $p_T$
and $Q \overline Q$ pair mass distributions are harder than the
photoproduction results in Figs.~\ref{gamArc}-\ref{gamAlb} since now
$\sqrt{S}$ is the full nucleon-nucleon center of mass energy.  In
hadroproduction, the quark rapidity distribution is symmetric around
$y_1=0$.

The $c \overline c$ $AA$ cross sections and rates are given in
Tables~\ref{AAccbar} and \ref{NAAccbar}.  The $b \overline b$
cross sections and rates are in Tables~\ref{AAbbbar} and
\ref{NAAbbbar} respectively.  Shadowing has less effect on the total
cross sections than in our previous calculations
\cite{spenprc,spenprl} which used earlier shadowing parameterizations
\cite{hpcshad,eskolanpb} with stronger gluon shadowing at low $x$ and
weaker gluon antishadowing.

Shadowing effects depend on the final state rapidity and pair mass.  The region
away from $y_1=0$ tends to be most sensitive to the shadowing
parameterization\cite{ekkv4,RVwz}.  At RHIC, the effect
grows with rapidity because at $y_1 \sim 0$ $x$ is not small and
shadowing is weak (charm) or $x$ is in the antishadowing region (bottom).  
Higher positive rapidities correspond to lower $x_2$ for the target (stronger
low $x$ shadowing) and larger $x_1$ for the projectile (the EMC
region), increasing the effect.

For $b\overline b$ production, the $x$ region moves from antishadowing
at $y_1=0$ to shadowing as $y_1$ increases.  In Au+Au collisions at
RHIC, $\sigma(S^i={\rm EKS98})/\sigma(S^i=1) = 1.22$ at $y_1=0$ and 0.917
at $y_1=2.5$.  At the LHC, the cross section varies less with rapidity
because both the target and projectile momentum fractions are in the
shadowing region.  In fact shadowing tends to decrease slightly with
rapidity because $S^i$ increases more with $x_1$ than it decreases
with $x_2$.

Nuclear shadowing is more important at the LHC than at RHIC because
the higher energy collisions probe smaller $x$ values.  Shadowing is
also larger for $c \overline c$ than $b \overline b$ because the
lighter charm quark is produced by lower $x$ partons than $b$ quarks.
Homogeneous shadowing reduces the $c \overline c$ total cross section
by $(3-5)$\% at RHIC while at the LHC the reduction is $(18-34)$\%.
As expected, the reduction is largest for the heaviest nuclei.  With
the inhomogeneous shadowing parameterization, EKS98$b$, the $c
\overline c$ cross section is reduced by $(1-2)$\% at RHIC and $\sim
10$\% at LHC.  The $b \overline b$ cross section is antishadowed at
RHIC by $(6-20)$\% using EKS98, reduced to $(3-8)$\% with EKS98$b$.
At the LHC, the $b \overline b$ cross section is reduced by
$(10-19)$\% with EKS98 and $\sim 6$\% with EKS98$b$.  
At large $b$, any inhomogeneity reduces the effect of shadowing.  The
larger the impact parameter cut,
the closer the peripheral $AA$ events mimic
hadroproduction in free proton collisions.

A direct comparison with the photoproduction cross sections in
Tables~\ref{gamAcc} and \ref{gamAbb} shows that the only rate
comparable to hadroproduction at RHIC is $c \overline c$
photoproduction in Au+Au collisions when the same impact parameter
range is considered.  The photoproduction cross section is $\sim 25$
mb, comparable to the 33 mb hadroproduction cross section.  With lead beams
at the LHC, the $c \overline c$ photoproduction cross section is
actually larger than the hadroproduction cross section.  The $b 
\overline b$ photoproduction cross section is always much lower 
than that of hadroproduction.

The hadroproduction cross sections tend to be larger than
photoproduction cross sections for several reasons.  
In hadroproduction, the full
$\sqrt{S}$ is available for heavy quark production while coherent
photon emission requires $\sqrt{S_{\gamma N}} \ll \sqrt{S}$.  The
lower energy reduces the cross section and restricts the $x$
range of the gluons taking part in the interactions.  Thus
hadroproduction probes smaller $x$ values than photoproduction.  At
low $x$ the gluon densities are larger than the photon flux.  

The minimum bias $pA$ results with $S^i = 1$ and $S^i$ = EKS98 are
presented in Tables~\ref{pAccbb} and \ref{NpAccbb} for charm and
bottom respectively.  We report the minimum bias $pA$ cross sections
only since it is difficult to select peripheral $pA$ events.
Shadowing is less important than in $AA$ collisions since the $pA$
cross section is linear in $S^i$ while the $AA$ cross section is
quadratic in $S^i$.  The minimum bias cross section is proportional to
$A$ for $S^i = 1$.  A comparison of the RHIC minimum bias $pA$ and the
peripheral $AA$ cross sections shows that the $pA$ cross section without
shadowing is equal to the $AA$ cross section divided by $(1/A)\int d^2b \,
T_{AA}(b > 2R_A)$, as expected.  There is no
corresponding factor of $A$ for photoproduction so the hadroproduction
$pA$ cross sections are always bigger than the photoproduction cross
sections in $pA$.  Recent studies have shown that a comparison of
hadroproduction in $pA$ and $pp$ collisions at the same energies can
provide detailed information on nuclear shadowing effects \cite{ekv}.
In this case, there is no difference in flux between $pA$ and
$AA$ collisions as there is in photoproduction.  Such studies could provide
important input to the $AA$ collisions discussed here.

The variations in the cross section due to quark mass and QCD scale 
are similar in both
hadroproduction and photoproduction.  However, the additional NLO
corrections are larger in hadroproduction.  Even at NLO the 
calculations do not always agree with data.  The measured
$B^+$ production rate in $p\overline p$ collisions at
$\sqrt{S}=1800$ GeV is $2.9\pm0.2\pm0.4$ times the NLO calculation
\cite{CDF2}.  The reason for this discrepancy is unknown but
some non-standard suggestions have been made \cite{harris,other}.
The discrepancy may also be due to an incomplete understanding of the
hadronization process \cite{matteo}.

The major uncertainty for hadroproduction in peripheral 
collisions is the minimum
impact parameter.  There is no known method for effectively selecting
very large-impact parameter hadronic events or, alternatively,
collisions with a small but well defined number of participants.  Zero
degree calorimeters (ZDCs) can be used to select events with a small
number of spectator neutrons but these come from a poorly defined
range of impact parameters.  For this reason, peripheral $AA$
collisions may best be studied by comparing different processes in a
ZDC-selected data set.  These different processes, such as heavy quark
and gauge boson production \cite{RVwz}, could be used to compare
nuclear parton densities for several species at a variety $x$ and
$Q^2$ values.

\section{Two-Photon Production}
\label{gamgamsec}

Heavy quark pairs can also be produced in purely electromagnetic
photon-photon collisions.  This process has been studied at $e^+e^-$
colliders.  However, in ion colliders the cross sections are enhanced
since the $\gamma \gamma$ luminosity increases as $Z^4$.

The $\gamma \gamma$ luminosity has been calculated by many authors
\cite{refs}.  Naively, it is given by the convolution of the photon
fluxes from two ultrarelativistic nuclei.  The photon flux from one
nucleus is the integral of $d^3N_\gamma/dk d^2r$ in Eq.~(\ref{wwr})
over $r$ excluding the nuclear interior.  Not all the flux is usable
because, when the nuclei actually collide, the two-photon interaction
products will be lost amongst the much denser hadronic debris.  The
usable flux is limited by the requirement that the nuclei do not
interact hadronically.  This is typically done by requiring that
$b>2R_A$.  However, when the photon energy is close to the kinematic
limit, $k_{\rm max} \approx \gamma_L \hbar c/R_A$, the flux is
sensitive to the exact choice of $R_A$.  To reduce the sensitivity to
$R_A$, we calculate the probability, $P(b)$, of a hadronic interaction
as a function of impact parameter,
\begin{equation}
P(b) = 1 - \exp\bigg[ -\sigma_{NN}^{\rm tot}(s) 
\int d^2r T_A(r)T_B(|\vec{b}-\vec{r}|)
\bigg] \, \, ,
\end{equation}
and use it to calculate the effective photon flux.
Woods-Saxon density distributions \cite{Vvv} are used to calculate the nuclear
overlap integral.  The nucleon-nucleon total cross section,
$\sigma_{NN}^{\rm tot}$, is 52 mb at 200 GeV and 93 mb at 5.5 TeV 
\cite{PDG}.  We use the total cross section to exclude any
interaction which could cause the nuclei to break up.  This soft
cutoff on the flux reduces the effective two-photon luminosity by a
few percent for $k \ll \gamma_L \hbar c/R_A$, rising to about
15\% at the kinematic limit compared to a hard cutoff with
$R_A=1.16(1-1.16A^{-2/3})A^{1/3}$ fm.
We also exclude the flux when the heavy quarks are produced inside
one of the nuclei.  Although these heavy quarks would probably
survive, the resulting interactions are likely to disrupt the nucleus,
making the collision appear hadronic.  With these exclusions, the
differential $\gamma \gamma$ luminosity is
\begin{equation}
{d{\cal L}_{\gamma\gamma} \over dk_1dk_2} = {\cal L}_{AA}
\int_{b>R_A}\int_{r>R_A} d^2b d^2r
\frac{d^3N_\gamma}{dk_1 d^2b}\frac{d^3N_\gamma}{dk_2 d^2r} 
P(|\vec b -\vec r|) \, \, .
\end{equation}

The two-photon center of mass energy, $\sqrt{S_{\gamma \gamma}}$, is
given by the photon energies, $S_{\gamma \gamma}=4k_1k_2$.  This
$S_{\gamma \gamma}$ is equivalent to $W^2$, commonly used in other
two-photon calculations.  The maximum $\sqrt{S_{\gamma\gamma}}$ is
$2k_{\rm max} = 2\gamma_L\hbar/R_A$, given in Table~\ref{gamfacs}.
This limit is a factor of $(\hbar c/m_pR_A)^2$ smaller than
$\sqrt{S}$, a factor of $10^{-3}$ for gold or lead.  Indeed, $2k_{\rm
max} < 2m_b$ for I+I and Au+Au collisions at RHIC.  Thus heavy quark
production in this channel should be much smaller than for photo- and
hadroproduction.

\begin{figure}
\setlength{\epsfxsize=0.95\textwidth}
\setlength{\epsfysize=0.5\textheight}
\centerline{\epsffile{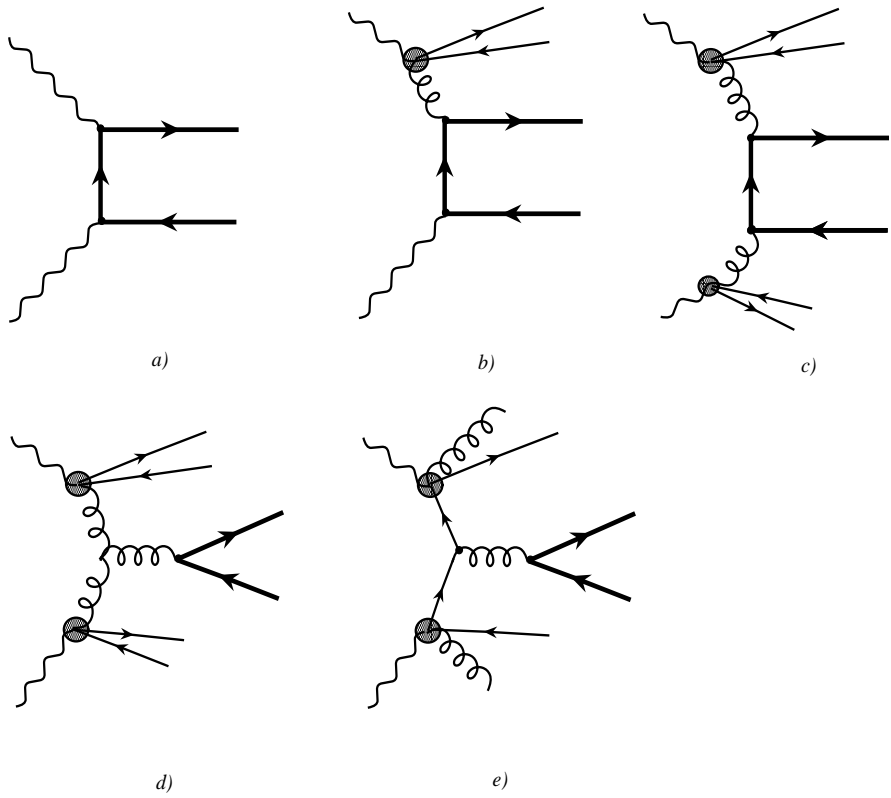}}
\caption[]{Feynman diagrams for two-photon production of heavy quarks
in (a) direct, (b) single-resolved, and (c)-(e) double-resolved
photons.  The crossed diagrams for (a) through (c) are not shown.}
\label{Feyn2phot}
\end{figure}

As in photoproduction, there are also direct and resolved
contributions.  Either one or both of the photons \cite{Drees} can
resolve itself into partons.  At the parton level, the single-resolved
photon case is equivalent to photoproduction while the double-resolved
photon situation is equivalent to hadroproduction.  Both of these
contributions are included here.  The diagrams for all of these
processes are shown in Fig.~\ref{Feyn2phot}.

The LO cross section for heavy quark production in direct
two-photon interactions is also proportional to $B_{\rm QED}$, as in 
Eq.~(\ref{gamApart}) for direct photoproduction, but with different couplings,
\begin{equation}
s^2 \frac{d^2 \sigma_{\gamma \gamma}}{dt_1 du_1} 
= 6 \pi \alpha^2 e_Q^4 B_{\rm
QED}(s,t_1,u_1) \delta(s + t_1 + u_1) \,\, , 
\label{gamgampart}
\end{equation}
where $s = (k_1 + k_2)^2 = S_{\gamma \gamma}$, $t_1 =
(k_1 - p_1)^2 - m_Q^2$, and $u_1 = (k_2 - p_1)^2 - m_Q^2$.  
Here $t_1 = U_1$ and $u_1 =
T_1$ for a detected quark.  The $\gamma \gamma \rightarrow Q \overline
Q$ cross section is a factor of $6 e_Q^2 \alpha/\alpha_s(Q^2)$ smaller than
the partonic $\gamma g \rightarrow Q \overline Q$ cross section,
Eq.~(\ref{gamApart}).  A 
factor of 3 comes from the three quark colors while another factor of 2 is 
from the spin averages.  The ratio
$\alpha/\alpha_s(Q^2)$ reduces the cross section for two-photon
production relative to photoproduction.  The same two Compton diagrams
apply to both two-photon production and photoproduction except that
a second photon replaces the gluon in $\gamma \gamma$ interactions.

The direct cross section is the convolution of the partonic two-photon 
cross section with
the two-photon luminosity for photon energies $k_1$ and $k_2$,
\begin{equation} 
S_{\gamma \gamma}^2 \frac{d^2 \sigma^{\rm dir}_{\gamma \gamma 
\rightarrow Q \overline Q}}{dT_1 dU_1}
= \int dk_1 dk_2 \frac{d
{\cal L}_{\gamma
\gamma}}{d k_1 d k_2} s^2 \frac{d^2 \sigma_{\gamma \gamma}}{dt_1 
du_1} \, \, .
\label{gamgamhad}
\end{equation}
The photon fluxes are exponentially suppressed for $k \geq \gamma_L
\hbar/R_A$.

The resolved processes follow from the discussion in Section~\ref{photosec}.
The cross section for singly resolved production of heavy quarks is
\begin{eqnarray}
S_{\gamma \gamma}^2  \frac{d^2 \sigma^{\rm 1-res}_{\gamma \gamma 
\rightarrow Q \overline Q}}{dT_1 dU_1}
= 2 \int dk_1 dk_2' \int \frac{dx_2}{x_2} \frac{d
{\cal L}_{\gamma
\gamma}}{d k_1 d k_2'} f_g^\gamma(x_2,Q^2) \hat{s}^2 
\frac{d^2 \sigma_{\gamma g}}{d \hat{t}_1 d \hat{u}_1} \, \, ,
\label{gam1reshad}
\end{eqnarray}
where the partonic invariants, $\hat{s} = (k_1 + x_2 k_2)^2 = x_2 
S_{\gamma \gamma}$,
$\hat{t}_1 = (k_1 - p_1)^2 - m_Q^2$, and $\hat{u}_1 = (x_2 k - p_1)^2
- m_Q^2$, are related to the total invariants by $\hat{t}_1 = U_1$ and 
$\hat{u}_1 = x_2T_1$ and $k_2' = x_2 k_2$.  The partonic cross section, 
$\sigma_{\gamma g}$, is
the photoproduction cross section in Eq.~(\ref{gamApart}).  The
cross section for double-resolved heavy quark production is
\begin{eqnarray}
S_{\gamma \gamma}^2  \frac{d^2 \sigma^{\rm 2-res}_{\gamma \gamma 
\rightarrow Q \overline Q}}{dT_1 dU_1}
& = & \int dk_1'  dk_2' \int \frac{dx_2}{x_2} 
\int \frac{dx_1}{x_1} \frac{d
{\cal L}_{\gamma
\gamma}}{d k_1' d k_2'} \bigg[ f_g^\gamma(x_1,Q^2)
f_g^\gamma(x_2,Q^2) \hat{ \hspace{-0.14em}\hat s}^2 
\frac{d^2 \sigma_{g g}}{d  \hat{ \hspace{-0.14em} \hat t}_1 
d \hat{ \hat u}_1 } \nonumber \\
& & \mbox{} + \sum_{i=u,d,s} ( f_q^\gamma (x_1,Q^2) f_{\overline
q}^\gamma (x_2,Q^2) + f_{\overline q}^\gamma (x_1,Q^2) f_q^\gamma(x_2,Q^2))
\hat{ \hspace{-0.14em}\hat s}^2 
\frac{d^2 \sigma_{q \overline q}}{d  \hat{ \hspace{-0.14em} \hat t}_1 
d \hat{ \hat u}_1} \bigg] \, \, ,
\label{gam2reshad}
\end{eqnarray}
where $k_2' = x_2 k_2$, $k_1' = x_1 k_1$ and the partonic invariants, $ \hat{
\hspace{-0.14em}\hat s} = (x_1 k_1 + x_2 k_2)^2 = x_1 x_2
S_{\gamma \gamma}$, $ \hat{ \hspace{-0.14em} \hat t}_1 = (x_1 k_1
- p_1)^2 - m_Q^2$, and $ \hat{ \hat u}_1 = (x_2 k_2 - p_1)^2 -
m_Q^2$, are related to the total invariants
by $ \hat{ \hspace{-0.14em} \hat t}_1 = x_1 U_1$ and $\hat{ \hat u}_1
= x_2 T_1$.  The partonic cross sections, $\sigma_{q \overline q}$ and
$\sigma_{gg}$, are given in Eqs.~(\ref{qqpartres}) and
(\ref{ggpartres}).  

The full two-photon heavy quark cross section is the sum of all three
contributions, 
\begin{eqnarray}
S_{\gamma \gamma}^2  \frac{d^2 \sigma_{\gamma \gamma
\rightarrow Q \overline Q}}{dT_1 dU_1} =
S_{\gamma \gamma}^2  \frac{d^2 \sigma^{\rm dir}_{\gamma \gamma
\rightarrow Q \overline Q}}{dT_1 dU_1} + 
S_{\gamma \gamma}^2  \frac{d^2 \sigma^{\rm 1-res}_{\gamma \gamma 
\rightarrow Q \overline Q}}{dT_1 dU_1} +
S_{\gamma \gamma}^2  \frac{d^2 \sigma^{\rm 2-res}_{\gamma \gamma
\rightarrow Q \overline Q}}{dT_1 dU_1} \, \, .
\label{gamgamtot}
\end{eqnarray}
In Eqs.~(\ref{gam1reshad}) and (\ref{gam2reshad}) the scale entering
into the photon parton densities and $\alpha_s(Q^2)$ has been set equal to 
$4m_Q^2$ due to the structure of the $\gamma \gamma$ luminosity.  
Changing the scale from $4m_T^2$ to $4m_Q^2$ 
increases the single-resolved cross section by
about 10\%, while the double-resolved cross section changes by at most 2\%.

\begin{figure}
\setlength{\epsfxsize=0.95\textwidth}
\setlength{\epsfysize=0.5\textheight}
\centerline{\epsffile{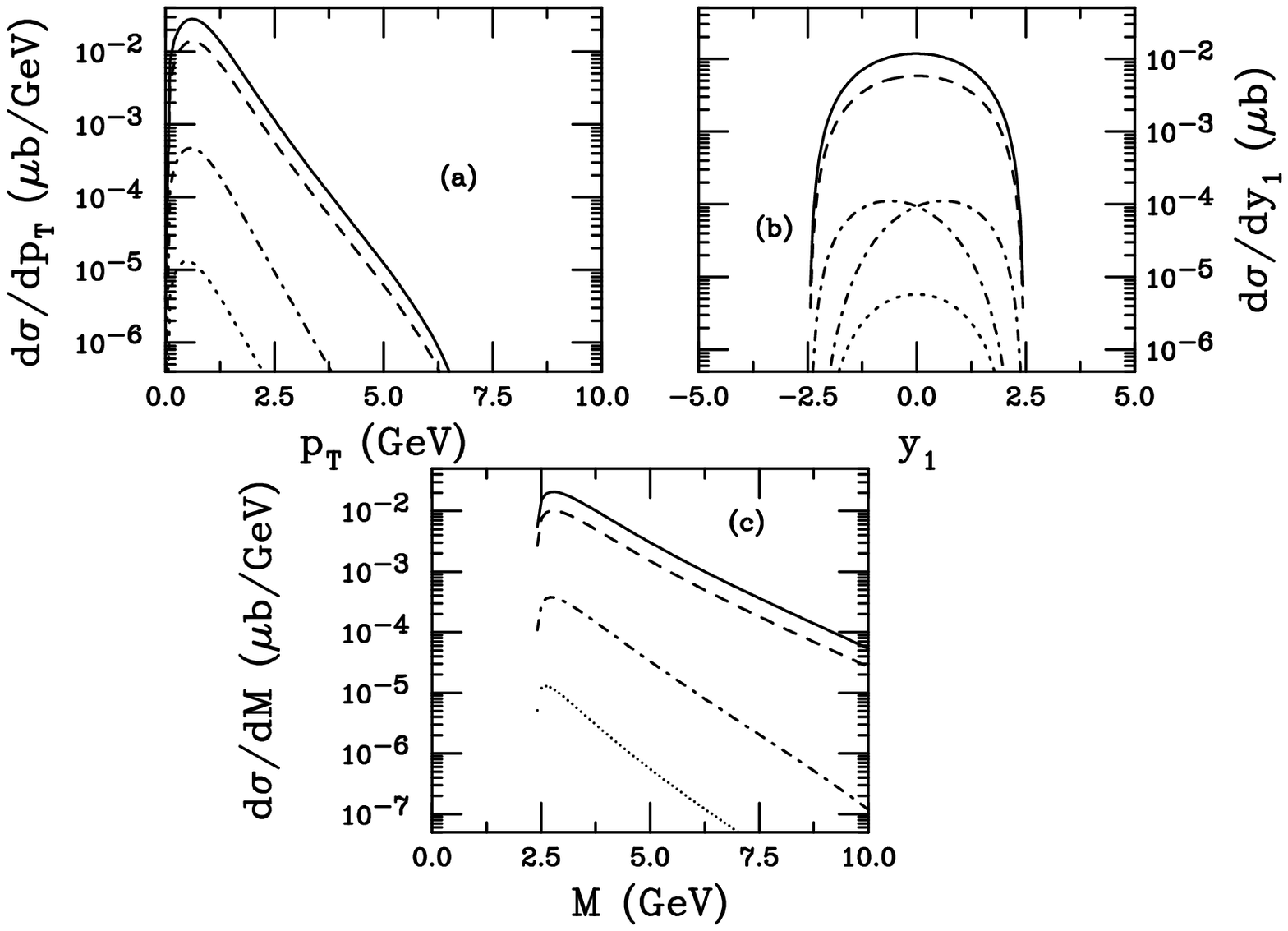}}
\caption[]{Charm production by two-photon processes in peripheral
Si+Si collisions at RHIC.  The results are shown for all pairs with no
mass cut.  The single $c$ quark $p_T$ (a) and rapidity (b)
distributions are shown along with the $c \overline c$ pair invariant
mass (c). The solid curve is the sum of all contributions: $\sigma^{\rm dir}$
(dashed), $\sigma^{\rm 1-res}$ (dot-dashed), and $\sigma^{\rm 2-res}$ 
(dotted).  The direct contribution is divided by two to facilitate comparison. 
Since either photon can be resolved, the single-resolved rapidity
distribution is reflected around $y_1=0$ to account for both sources.}
\label{ggrc}
\end{figure}

\begin{figure}
\setlength{\epsfxsize=0.95\textwidth}
\setlength{\epsfysize=0.5\textheight}
\centerline{\epsffile{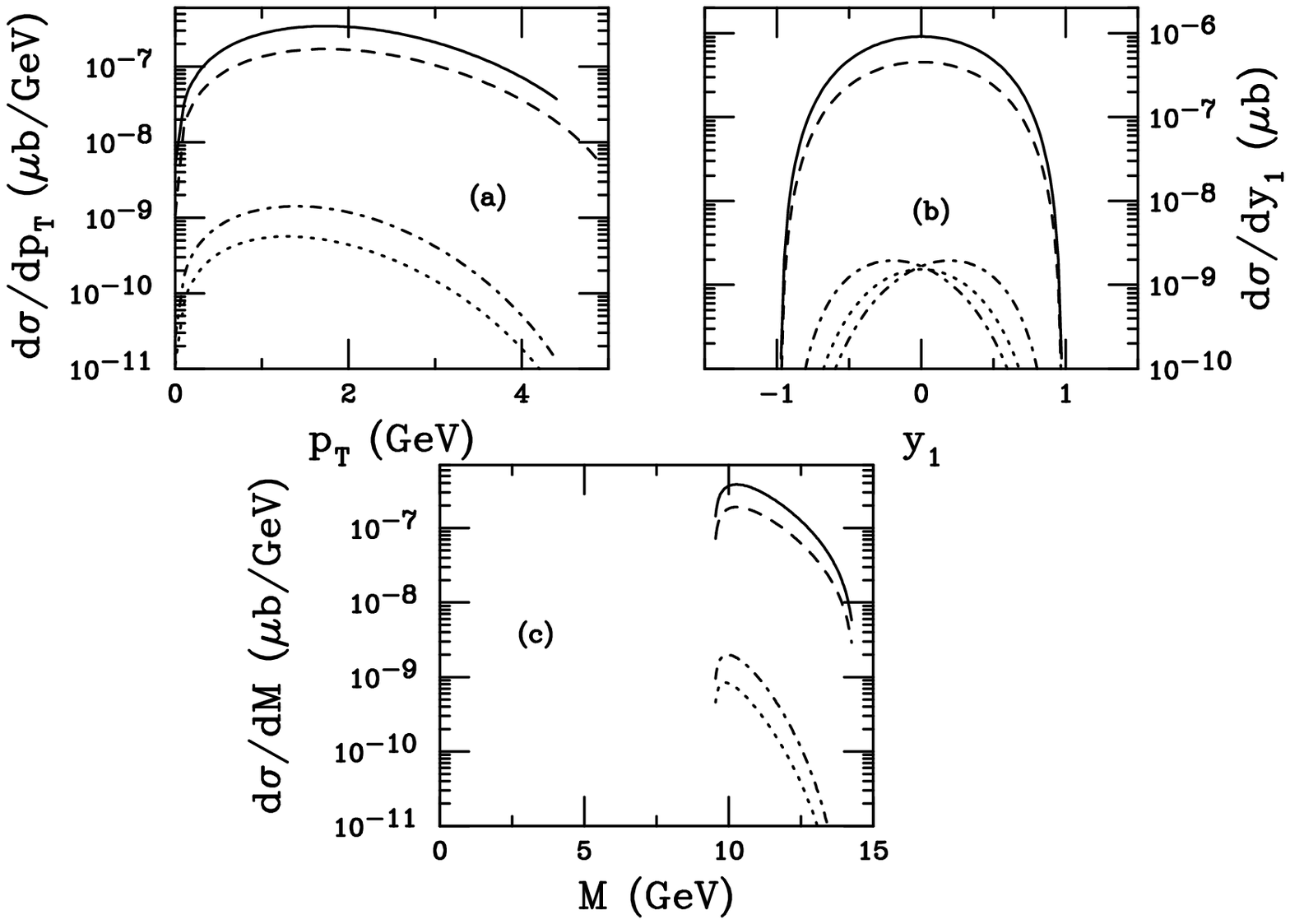}}
\caption[]{Bottom production by two-photon processes in peripheral
Si+Si collisions at RHIC.  The results are shown for all pairs with no
mass cut.  The single $b$ quark $p_T$ (a) and rapidity (b)
distributions are shown along with the $b \overline b$ pair invariant
mass (c). The solid curve is the sum of all contributions: $\sigma^{\rm dir}$
(dashed), $\sigma^{\rm 1-res}$ (dot-dashed), and $\sigma^{\rm 2-res}$ 
(dotted).  The direct contribution is divided by two to facilitate comparison. 
Since either photon can be resolved, the single-resolved rapidity
distribution is reflected around $y_1=0$ to account for both sources.}
\label{ggrb}
\end{figure}

\begin{figure}
\setlength{\epsfxsize=0.95\textwidth}
\setlength{\epsfysize=0.5\textheight}
\centerline{\epsffile{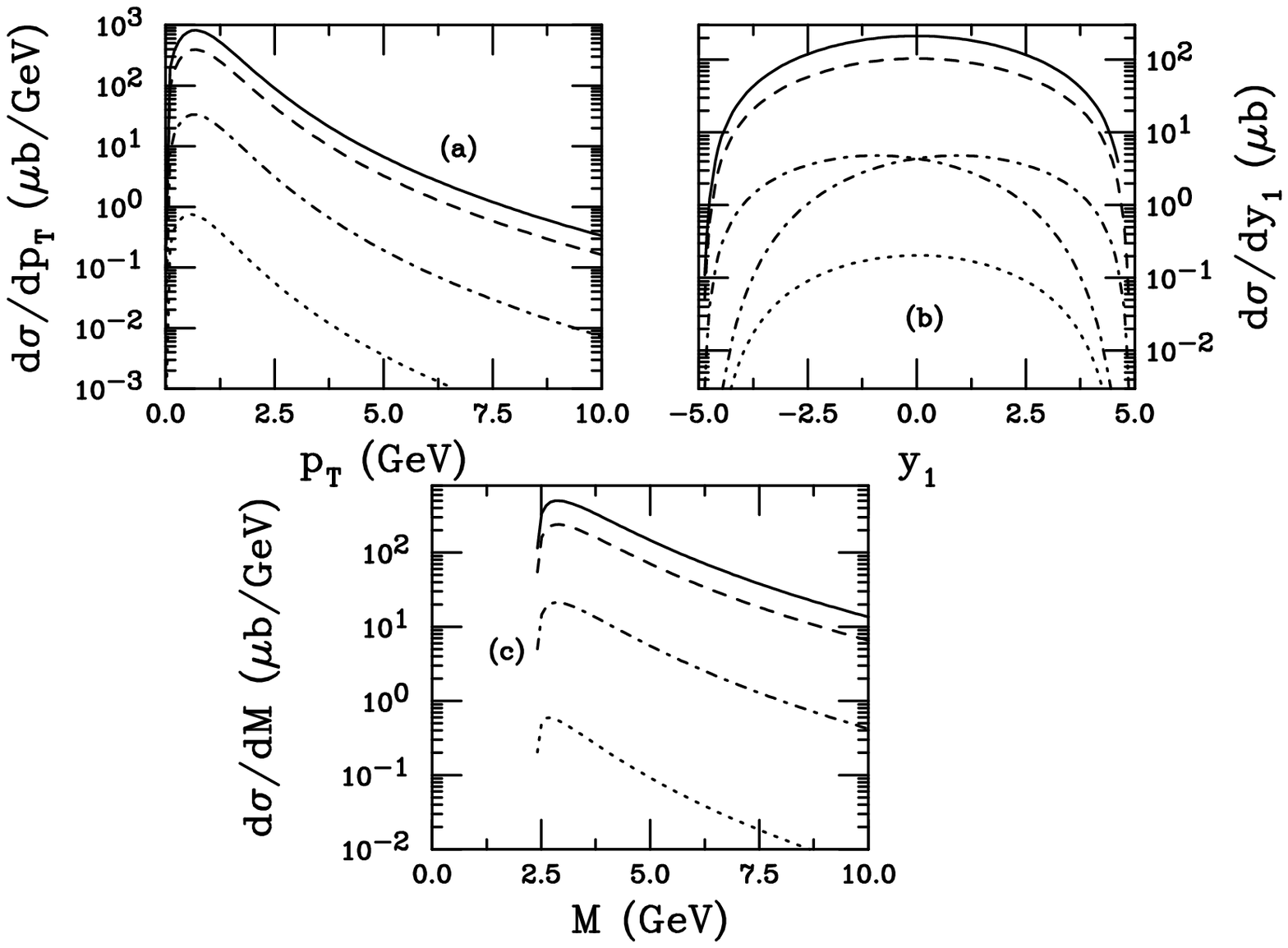}}
\caption[]{Charm production by two-photon processes in peripheral
Pb+Pb collisions at LHC.  The results are shown for all pairs with no
mass cut.  The single $c$ quark $p_T$ (a) and rapidity (b)
distributions are shown along with the $c \overline c$ pair invariant
mass (c). The solid curve is the sum of all contributions: $\sigma^{\rm dir}$
(dashed), $\sigma^{\rm 1-res}$ (dot-dashed), and $\sigma^{\rm 2-res}$ 
(dotted).  The direct contribution is divided by two to facilitate comparison. 
Since either photon can be resolved, the single-resolved rapidity
distribution is reflected around $y_1=0$ to account for both sources.}
\label{gglc}
\end{figure}

\begin{figure}[p]
\setlength{\epsfxsize=0.95\textwidth}
\setlength{\epsfysize=0.5\textheight}
\centerline{\epsffile{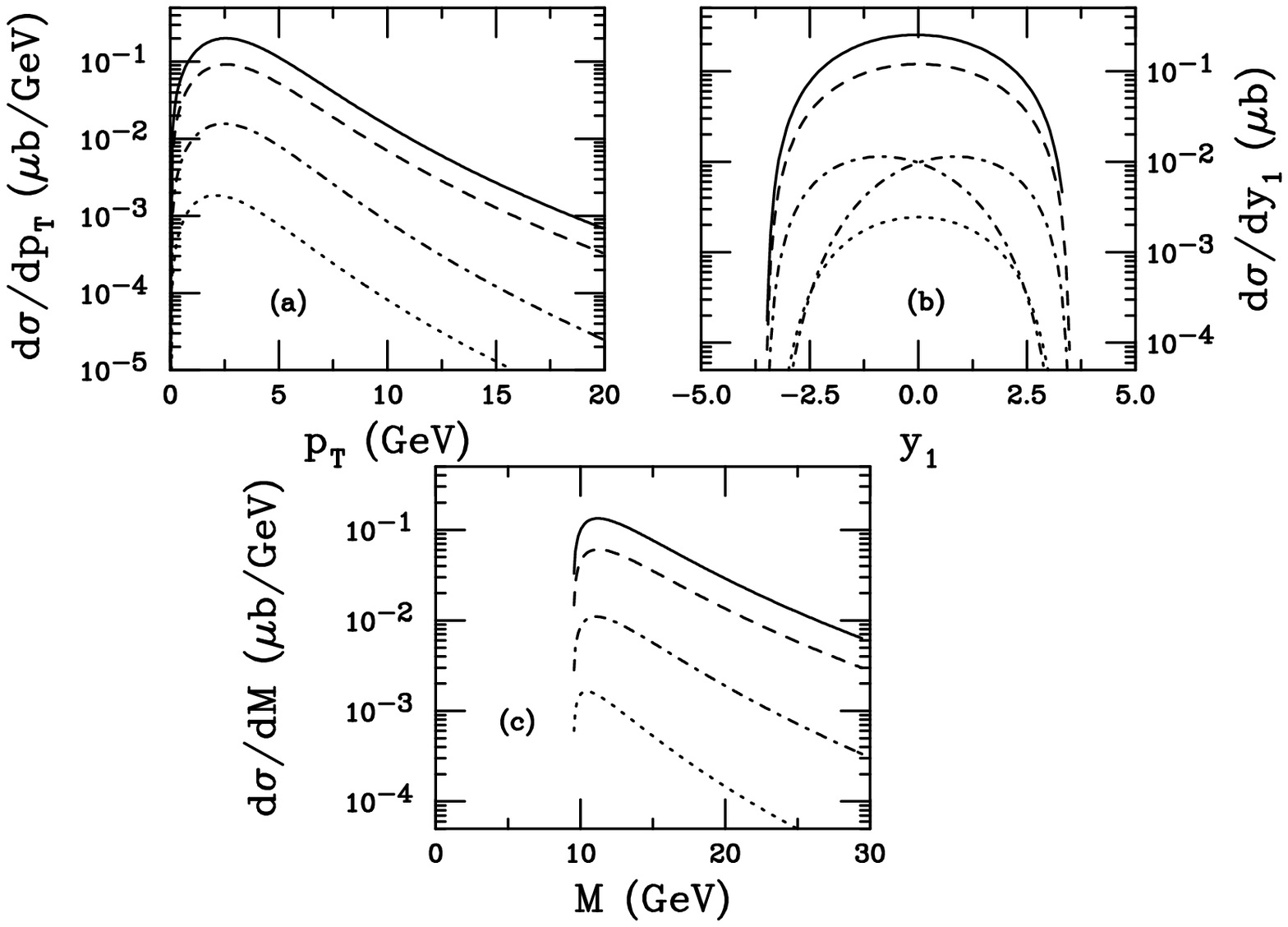}}
\caption[]{Bottom production by two-photon processes in peripheral
Pb+Pb collisions at LHC.  The results are shown for all pairs with no
mass cut.  The single $b$ quark $p_T$ (a) and rapidity (b)
distributions are shown along with the $b \overline b$ pair invariant
mass (c). The solid curve is the sum of all contributions: $\sigma^{\rm dir}$
(dashed), $\sigma^{\rm 1-res}$ (dot-dashed), and $\sigma^{\rm 2-res}$ 
(dotted).  The direct contribution is divided by two to facilitate comparison. 
Since either photon can be resolved, the single-resolved rapidity
distribution is reflected around $y_1=0$ to account for both sources.}
\label{gglb}
\end{figure}

Figures~\ref{ggrc}-\ref{gglb} show the corresponding $c \overline c$
and $b \overline b$ production distributions. The RHIC results are
shown for Si+Si collisions since that is the largest $A$ for which
$2k_{\rm max} > 2m_b$.  The Pb+Pb results are shown for
LHC.  The integrated cross sections for all the other nuclei are given in
Tables~\ref{gamgamcc} and \ref{gamgambb}.  The quark $p_T$ and
rapidity and the $Q \overline Q$ pair invariant mass distributions are
narrower for the heavier nuclei due to the lower associated
$\sqrt{S_{\gamma \gamma}}$.  The production is mostly direct.  The single and
double resolved $p_T$ and mass distributions are narrower than the
direct results at all energies.

The rapidity distributions are symmetric around $y_1=0$ except for the
singly resolved processes.  Since either photon may be resolved, we give
the single-resolved rapidity distributions in both cases. The total
single-resolved rapidity distribution is the sum.  We present
both to be consistent with the direct and double-resolved
calculations.  The factor of two in the cross section, 
Eq.~(\ref{gam1reshad}), is included in the single-resolved
transverse momentum and invariant mass distributions.

Direct $\gamma \gamma$ production dominates two-photon production of
heavy quarks.  For $c \overline c$ pairs, at RHIC the single-resolved
cross section is $(0.6-1.6)$\% of the direct production.  Double-resolved
production is $(2.5-3.3)$\% of the single resolved.  
The single-resolved to direct ratio increases with
$A$ or, equivalently, decreasing $\sqrt{S_{\gamma
\gamma}}$.  Interestingly, the single- to double-resolved cross section
ratios decrease with increasing $A$,
presumably due to the larger $q \overline q$ component at lower
energies, closer to production threshold where the quark distribution
in the photon is dominant.  At LHC, the single-resolved cross section
is 4.5\% of the direct production.  The single-resolved component is higher
at LHC because of the higher energy.  However, double-resolved production is
only 2.1\% of the single resolved because the higher beam energy
reduces $x$, increasing the photon gluon distribution but the $gg$ 
contribution is too small to make up the difference.  The situation
changes for $b \overline b$ production.  At RHIC, single-resolved
production is $(0.3-0.7)$\% of the direct component but double-resolved
production is 20\% of the single resolved.  The lower energy strongly 
reduces the single-resolved cross section relative to the
direct $\gamma \gamma$ but has less effect on double-resolved
production because of the strong $q \overline q$ component.  At the
LHC the single-resolved result is 10\% of the direct while the
double-resolved is reduced relative the single-resolved by the same
factor.  The LEP results suggest relatively low quark masses and a
large resolved cross section \cite{photon01}.  Our resolved contributions are
smaller, possibly because the $\gamma \gamma$ luminosity at LEP drops
more slowly with $\sqrt{S_{\gamma \gamma}}$ than do the $\gamma
\gamma$ luminosities for ions.

These results are for all $Q \overline Q$ pairs, as are the results
shown for the other channels.  However, a significant fraction of the
pairs have masses below the $H \overline H$ threshold, especially at
RHIC.  Pairs with mass $M<2m_H$ are also produced in photo- and
hadroproduction.  A few of these pairs will become quarkonium states
\cite{hpcpsi}.  Most of them will hadronize into heavy-flavor hadrons,
thanks to the presence of accompanying particles.  A soft gluon can
provide the energy to bring the quarks on shell.  However, two-photon
interactions occur in a vacuum with no available energy source.
Pairs with $M<2m_H$ may annihilate if they do not form quarkonium.
Tables~\ref{gamgamcc} and \ref{gamgambb} compare the total cross
sections for all $Q \overline Q$ pairs to those pairs with $M>2m_H$.
In both cases, the two-photon cross sections are several orders of
magnitude below those for hadroproduction and photoproduction.  The $c
\overline c$ cross sections are ${\cal O}$(nb) rather than mb and the
$b \overline b$ cross sections are ${\cal O}$(pb) rather than $\mu$b.

The requirement $M>2m_H$ dramatically reduces charm production.  At
RHIC the $c \overline c$ cross section is reduced a factor of $3-6$ for
direct $\gamma \gamma$, $4-16$ for single resolved, and $5-22$ for the
double-resolved.  The higher LHC energy lessens the threshold effect
considerably; the reduction is only a factor of $\sim 1.9$ for
direct and single-resolved production and $\sim 2.5$ for
double-resolved production.  These reductions depend strongly on the
heavy quark mass.  Our $m_c$, 1.2 GeV, is about $0.64 \, m_D$ but the
reduction is much smaller for larger charm masses.  This
threshold effect reduces the overall sensitivity of the calculation to
the charm quark mass.  Charm production calculations with a threshold
cut match recent LEP data\cite{Andreev01}, indicating that the
reduced cross sections are more appropriate experimentally.

The threshold reduction is smaller for bottom quarks since $m_b = 4.75 \,
{\rm GeV} \, 
\approx 0.9 m_B$.  At RHIC, the cross section is a factor of $\sim
1.5$ smaller for direct photoproduction and $1.5-2.5$ for single- and
double-resolved production.  At the LHC, all the cross sections are
reduced by $(10-20)$\%.   The threshold effect is more important
for larger $A$ because $\sqrt{S_{\gamma \gamma}}$ falls with
increasing $R_A$.

\begin{figure}[p]
\setlength{\epsfxsize=0.95\textwidth}
\setlength{\epsfysize=0.25\textheight}
\centerline{\epsffile{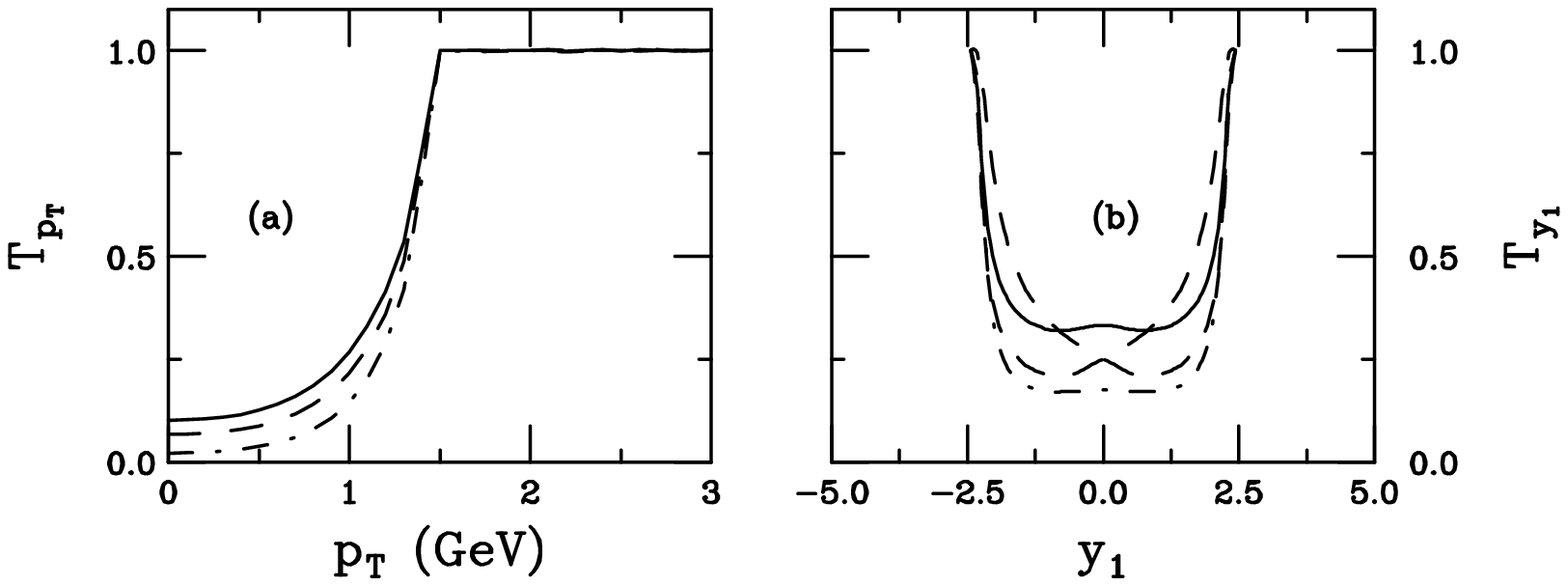}}
\caption[]{Reduction in charm production due to the requirement that
$M>2m_D$ in two-photon production in
peripheral Si+Si collisions at RHIC.  The ratio of the cross section
above threshold relative to the total cross section is shown as a
function of $p_T$ (a) and $y_1$ (b) for $\sigma^{\rm dir}$ (solid),
$\sigma^{\rm 1-res}$ (dashed) and $\sigma^{\rm 2-res}$ (dot-dashed).  Since
either photon can be resolved, the single-resolved rapidity ratio is
reflected around
$y_1=0$ to account for both sources.}
\label{ggrcsub}
\end{figure}

\begin{figure}[p]
\setlength{\epsfxsize=0.95\textwidth}
\setlength{\epsfysize=0.25\textheight}
\centerline{\epsffile{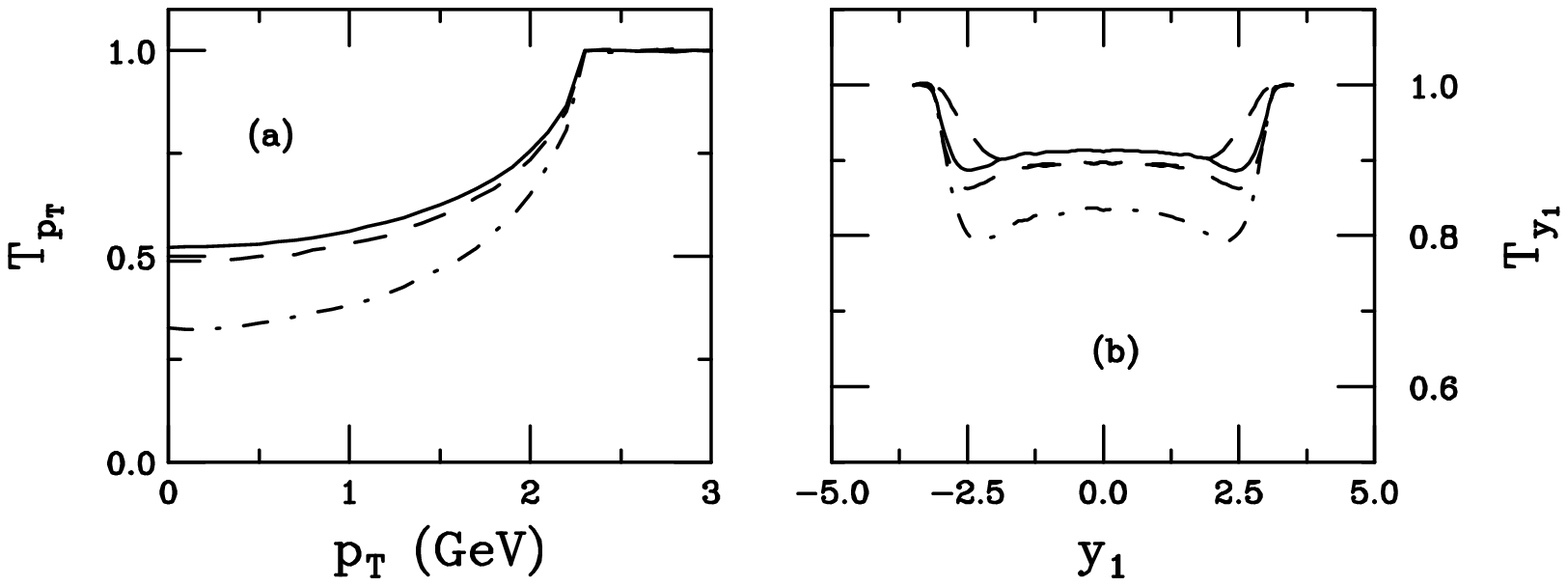}}
\caption[]{Reduction in bottom production due to the requirement that
$M>2m_B$ in two-photon production in
peripheral Pb+Pb collisions at LHC.  The ratio of the cross section
above threshold relative to the total cross section is shown as a
function of $p_T$ (a) and $y_1$ (b) for $\sigma^{\rm dir}$ (solid),
$\sigma^{\rm 1-res}$ (dashed) and $\sigma^{\rm 2-res}$ (dot-dashed).  Since
either photon can be resolved, the single-resolved rapidity ratio is
reflected around
$y_1=0$ to account for both sources.}
\label{gglbsub}
\end{figure}

Figures~\ref{ggrcsub} and~\ref{gglbsub} show the ratios for $Q
\overline Q$ production with and without the threshold cut as a
function of $p_T$ and rapidity for $c \overline c$ production at RHIC
and $b\overline b$ production at the LHC. An invariant mass cut simply
selects pairs with $M > 2m_H$.  The threshold cut only affects low
$p_T$ quarks. The minimum $p_T$ for a quark to pass the $2m_H$
threshold is $p_T \geq \sqrt{m_H^2 - m_Q^2}$, 1.4 GeV for $m_c = 1.2$
GeV and 2.3 GeV for $m_b = 4.75$ GeV.  Naturally, with a larger quark
mass, the minimum $p_T$ would decrease.  The larger threshold effect
for charm production appears because the peak of the $p_T$
distributions in Figs.~\ref{ggrc} and \ref{gglc} is below this minimum
$p_T$.  The average $p_T$ for bottom production is larger so that more
of the cross section survives the threshold cut.  On the other hand,
the cross section is reduced most near $y_1=0$ where low $p_T$
dominates the rapidity distribution.  At large rapidities, the pair
$Q\overline Q$ mass is always above threshold.  The threshold has the
smallest effect on direct production and the strongest on the
double-resolved cross sections, as can be expected due to the
decreasing effective energy available for each process.

Energy conservation requires that a heavy quark pair produced with
mass $M$ retain that energy.  To compensate for the `mass deficit',
$M-2m_H$, the final state mesons must have less kinetic energy than
the initial state quarks.  Near threshold, the quark
$p_T$ and rapidity distributions presented in Figs.~\ref{ggrc}-\ref{gglb} 
will differ from the
final state meson distributions.

We do not present any two-photon results for $pA$ since the cross
sections are very small and
the proton substructure could play a role \cite{ohnemus}.  On the
other hand, the small proton radius allows $pA$ collisions to reach
higher $\gamma\gamma$ energies than the corresponding $AA$ collisions so that
$b\overline b$ production would be energetically accessible in $p$I and $p$Au
collisions at RHIC.

The major uncertainties in the $\gamma\gamma$ calculations are the
quark masses and the $\gamma\gamma$ luminosity.  In contrast to
hadroproduction and photoproduction, the higher order corrections seem
fairly well known.  If the $\gamma\gamma$ luminosity can be
sufficiently well determined, heavy quark production could then be used to fix
the quark masses.  The uncertainties in the $\gamma\gamma$
luminosity are comparable to those for $\gamma A$ and also depend on
the minimum impact parameter.  However, final states with known
$\gamma\gamma$ couplings can be used for calibrations.  Lepton pair
production covers the full range of $S_{\gamma\gamma}$ and may be
accurately calculated using only electrodynamics.  Production of
well known mesons may also help check the luminosity.  With these
calibrations, $\sigma(\gamma\gamma\rightarrow Q\overline Q)$ could
clearly be measured to the 10\% level.  At that point, other theoretical
uncertainties will dominate and the measurements can be used to
determine the heavy quark masses.  These masses can then be used in 
calculations of other processes.

\section{Theoretical Comparisons}
\label{thsec}

In this section, we compare and contrast some of the calculational
uncertainties in our results.  We have performed fully LO
calculations, including LO parton densities and a one-loop evaluation
of $\alpha_s(Q^2)$. We first compare our LO results with NLO calculations.
We also describe the dependence of our results on the chosen quark
mass and scale.

At any order, the partonic cross sections for all three processes
studied may be expressed in terms of dimensionless scaling functions
$f^{(k,l)}_{ij}$ that depend only on the variable $\eta = \hat s/4 m^2
- 1$ \cite{KLMV},
\begin{eqnarray}
\label{scalingfunctions}
\hat \sigma_{ij}(\hat s,m_Q^2,Q^2) = \frac{(\alpha e_Q^2)^q
\alpha^p_s(Q^2)}{m_Q^2}
\sum\limits_{k=0}^{\infty} \,\, \left( 4 \pi \alpha_s(Q^2) \right)^k
\sum\limits_{l=0}^k \,\, f^{(k,l)}_{ij}(\eta) \,\,
\ln^l\left(\frac{Q^2}{m_Q^2}\right) \, , 
\end{eqnarray} 
where $\hat s$ is the square of the partonic center of mass energy,
$m_Q$ is the heavy quark mass, and $Q^2$ is the scale.  The cross sections
are expanded in powers of $\alpha_s(Q^2)$ and $\alpha$.  The exponents $q$
and $p$ depend on the initial process: $q = 1$, $p=1$ direct
photoproduction; $q = 0$, $p=2$ hadroproduction; and $q=2$, $p=0$
direct two-photon production.  The summation over $k$ includes all
corrections beyond LO which only involve powers of $\alpha_s(Q^2)$ with
$k=0$ corresponding to the Born and $k=1$ to the NLO cross sections.
It is only at NLO that the logarithms $\ln^l(Q^2/m_Q^2)$ appear.  Two
scales, for renormalization and factorization, appear in the
calculation but they are generally assumed to be the same since this
choice is used in fits of the parton densities.  The total cross
sections are obtained by convoluting the partonic cross sections with
the parton distribution functions or photon fluxes.  The scale $Q^2$
enters the total cross section in the evaluation of $\alpha_s(Q^2)$ and in
the parton densities of the proton or photon (for resolved processes).

For a fully consistent calculation of the partonic cross sections,
$\alpha_s(Q^2)$ should be evaluated to one loop when $k=0$, two loops when
$k=1 \cdots$.  The strong coupling constant has been evaluated up to
three loops, corresponding to $k=2$.  However, a consistent evaluation
of the cross section, order-by-order in partonic cross section, parton
distribution, and $\alpha_s(Q^2)$, is usually not done.  One is
usually interested only in the effect of the next-higher-order
corrections to the total cross section and it is only possible to
measure the change by leaving other inputs, such as the parton
densities, the same at all orders.  Thus, theoretical ratios of the
total NLO to LO cross sections, the $K$ factors, are typically
independent of the observable \cite{RVzphys,RVinprog}.

The hadroproduction $K$ factor is larger for `light' heavy quarks, low $p_T$,
and low $M$.  As the heavy quark mass increases, $K$ drops from
$2.5-3$ for $c \overline c$ to $1.8-2.3$ for $b \overline b$ in this
energy range.  For direct photoproduction, the $K$
factors are smaller.  The calculated $K$ factors for direct
photoproduction of bottom are $1.4-1.7$ for $\sqrt{S_{\gamma p}} =
314$ and 1265 GeV respectively \cite{SvN}.  The LO resolved
photoproduction results, ${\cal O}(\alpha_s^2)$ with $p=2$, $q=0$ in
Eq.~(\ref{scalingfunctions}) as at LO in hadroproduction, are
typically used without NLO corrections in photoproduction so that the
same order in $\alpha_s(Q^2)$ is used for both direct and resolved
photoproduction.  Thus the $K$ factor would only be constant with
rapidity and transverse momentum for direct photoproduction, not for
the sum in Eq.~(\ref{phottot}).  However, the resolved contribution is
always rather small and should not significantly change the $K$
factor.  The NLO $\gamma\gamma$ corrections are smaller still, $K =
1.33$ for $c \overline c$ and 1.24 for $b \overline b$ in direct
$\gamma \gamma$, dropping to $K \sim 1.15$ for $c \overline c$ and 1.21 for
$b \overline b$ single-resolved production \cite{Drees}.  The small
$K$ factor for direct $\gamma \gamma$ should perhaps not be a surprise
since, in a sense, $\gamma \gamma$ is the cleanest determination of
the $K$ factor because there is no parton density, $\alpha_s(Q^2)$ or scale
dependence at LO.

This $K$ factor, calculated with both the LO and NLO scaling functions
convoluted with NLO parton densities and two-loop evaluations of
$\alpha_s(Q^2)$, does not mesh with a full LO calculation using LO parton
densities.  The incompatibilities include the difference in $\alpha_s(Q^2)$
evaluated at one and two loops.  In the MRST LO densities, $\Lambda_3 = 0.204$
GeV so that $\alpha_s^{\rm 1-loop} = 0.287$ and
$\alpha_s^{\rm 2-loops} = 0.220$.  The hadronic LO cross sections
calculated with the MRST HO distributions\cite{pdf} are $\sigma_{\rm LO}
= 196$ $\mu$b at 200 GeV, rising to 6.03 mb at 5.5
TeV, compared with $\sigma_{\rm NLO} = 382$ $\mu$b at 200 GeV and 5.83 mb
at 5.5 TeV.  The NLO evaluation is two times larger at RHIC, but at
the LHC, the results agree within 3\%.  The difference is mostly due
to the higher one-loop $\alpha_s(Q^2)$.  Because of these variations, we do
not apply any $K$ factors to our LO calculations.

Our calculations for all three processes use the same values of $m_Q$
and $Q^2$.  The values are chosen to agree with hadroproduction data
at fixed target energies. Photoproduction and two-photon data imply 
larger charm quark masses than used here.  The typical charm mass used
for those processes, $1.6-1.7$ GeV, predict lower cross sections than
those employing
the quark masses obtained from hadroproduction.  One can speculate as
to why this is true. It may be that the incident quarks and gluons in
hadroproduction interact with the virtual heavy quark at its current
quark mass while the almost real photons interact with the constituent
$c$ and $b$ quarks.  The constituent quark mass is larger than the
current quark mass and is thus more compatible with the photoproduction
data.  On the other hand, since $K>2$ for hadroproduction, 
unincorporated higher order corrections may
explain the apparent need for different quark masses.  Near threshold,
the $b \overline b$ cross section has been evaluated to
next-to-next-to-leading order and next-to-next-to-leading logarithm
(NNLO-NNLL).  Recent results from HERA-B \cite{HERA-Bbb} agree very
well with the predicted $30 \pm 8 \pm 10$ nb NNLO-NNLL cross section
\cite{KLMV} calculated with $m_b = 4.75$ GeV.  The NLO evaluation at
the same energy is a factor of two smaller, suggesting that 
NLO calculations require smaller bottom quark masses.
However, the NNLO-NNLL expansion is invalid far away from threshold so
that the importance of further higher order corrections is difficult
to quantify.

\begin{figure}
\setlength{\epsfxsize=0.95\textwidth}
\setlength{\epsfysize=0.6\textheight}
\centerline{\epsffile{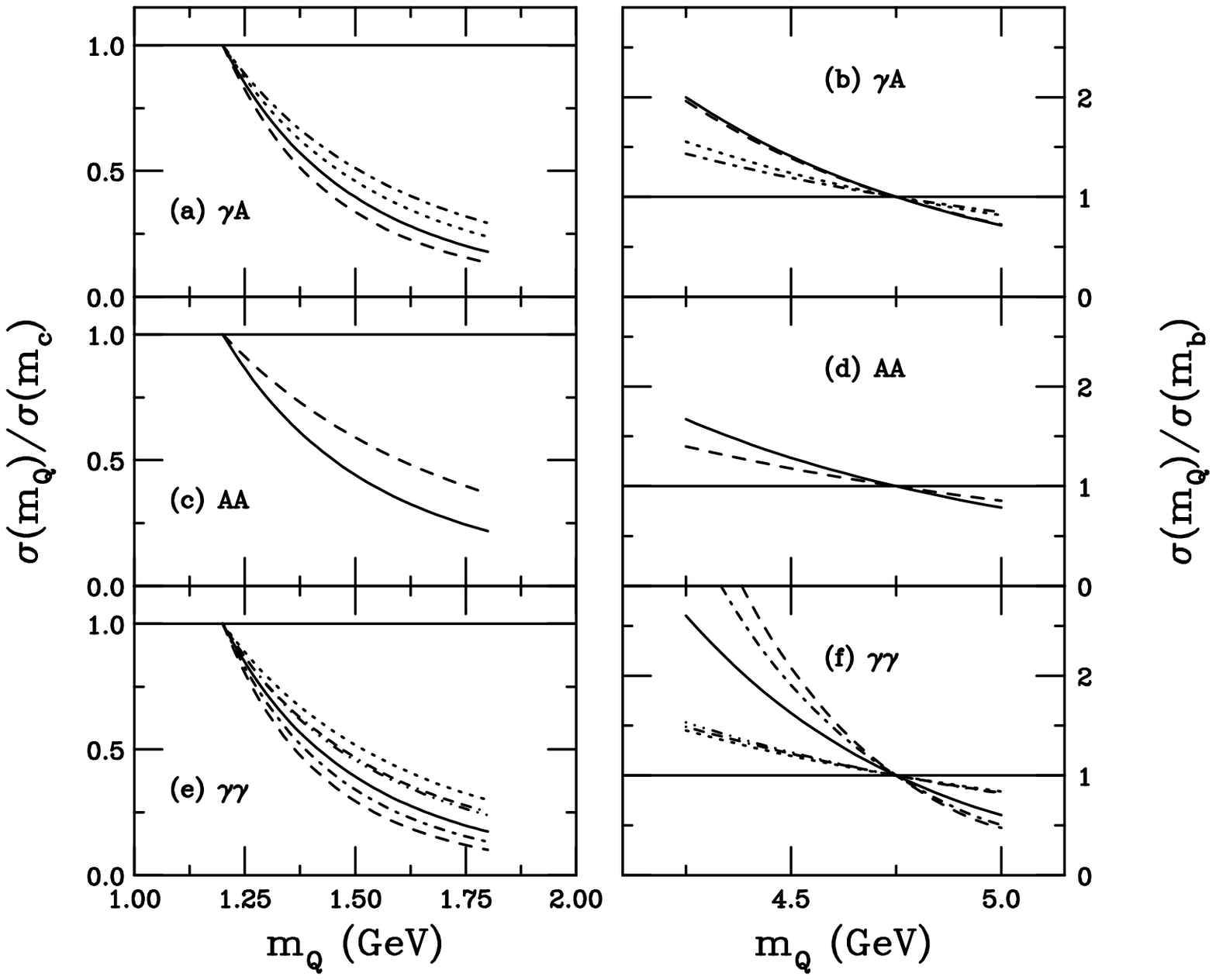}}
\caption[]{The quark-mass dependence of our calculated cross sections.
The left side is for charm, normalized to the cross sections with
$m_c=1.2$ GeV while the right side is for bottom, normalized to cross
sections with $m_b=4.75$ GeV.  The RHIC results are for Si+Si
interactions while the LHC results are for Pb+Pb interactions.  The
photoproduction ratios in (a) and (b) are for direct (solid--RHIC;
dot-dashed--LHC) and resolved (dashed--RHIC; dotted--LHC) production.
The hadroproduction results at RHIC and LHC are given by the solid and
dashed curves respectively in (c) and (d).  The two-photon ratios in
(e) and (f) are for $\sigma^{\rm dir}$ (solid--RHIC; dotted--LHC),
$\sigma^{\rm 1-res}$ (dashed--RHIC; dot-dot-dot-dashed--LHC) and
$\sigma^{\rm 2-res}$ (dot-dashed--RHIC; dot-dash-dash-dashed--LHC).}
\label{massdep}
\end{figure}

Figure~\ref{massdep} shows the quark mass dependence of the total
cross sections for all three processes.  
We plot $\sigma(m_Q)/\sigma(m_0)$ where $m_0 =
1.2$ GeV for charm and 4.75 GeV for bottom.  The scale used is $Q^2
\propto 4m_c^2$ and $m_b^2$ for charm and bottom, respectively.  Results
are shown for $\sqrt{S} = 250$ GeV Si+Si collisions at RHIC and 5.5 TeV
Pb+Pb collisions at the LHC.  The mass sensitivity 
is smaller at higher
energies, as expected.  For a given energy, hadroproduction is the
least mass dependent.  The direct photoproduction and two-photon
production processes have nearly the same mass dependence.  Resolved
production has a stronger mass dependence, especially at RHIC.  The
mass dependence is stronger for charm than bottom, mainly because the
charm mass is varied by 50\%, from 1.2 to 1.8 GeV while the bottom mass is
only varied 18\%, from 4.25 to 5.00 GeV.

The photoproduction and two-photon cross sections are more mass
dependent than hadroproduction at the same energies.  The maximum
$\gamma p$ collision energies, $\sqrt{S_{\rm max}}$, are a factor of
$4-6$ less than $\sqrt{S}$ in $AA$ collisions at both colliders.  The
maximum photon energy fraction, $\hbar c/m_pR_A$, is $\sim 0.03$ for
gold or lead, rising to 0.1 for silicon, much smaller than the maximum
parton energy fraction, $x=1$.  This energy deficit is difficult to
overcome and is only compensated for by the photon flux factor of
$Z^2$ in charm production with the heaviest ions.  The energy
difference is more important for bottom production, especially at RHIC
since $\sqrt{S_{\rm max}}$ is close to the $b \overline b$ threshold.
As Table~\ref{lum} shows, lighter ions may be advantageous for
photoproduction studies since the higher photon energies and
accelerator luminosities can compensate for the smaller cross
sections.  The energy deficit in $\gamma \gamma$ production is even
larger, $9.8-31.25$ between the maximum $\sqrt{S_{\gamma \gamma}}$ and
$\sqrt{S}$.  For the heaviest ions at RHIC, the maximum
$\sqrt{S_{\gamma \gamma}}$ is below the $b \overline b$ threshold.  In
these collisions, even charm is not far from threshold.  The factor of
$Z^4$ cannot compensate for such an energy deficit.  The 1 mb $c
\overline c$ cross section for $M>2m_c$ at LHC is still a factor of
1000 lower than those of the other processes.  Thus good experimental
separation is essential for observing clean $\gamma \gamma$
interactions.

The photo- and hadroproduction scale dependence is small.  The cross
sections only change a few percent between $Q^2 = m_Q^2$ and $4m_Q^2$
because increasing the scale decreases $\alpha_s(Q^2)$ but increases the
gluon density $F_g^p(x,Q^2)$ and vice versa.  The two effects largely
cancel.  At NLO, the scale dependence is usually larger for charm and
bottom quarks than at LO \cite{hpc} because $\alpha_s(Q^2)$ multiplies the
logarithm $\ln(Q^2/m_Q^2)$.  The scale dependence only enters through the
resolved contributions in $\gamma \gamma$ production where the effect
is a factor of $1.5-2$ on single-resolved production and $1.05-1.4$ on
double-resolved production.  However, the total cross section is essentially
unaffected because the direct contribution is independent of scale.

\section{Experimental Separation}
\label{exsec}

To study ultra-peripheral heavy quark production, it is necessary to
be able to disentangle the three channels.  Photoproduction,
hadroproduction, and two-photon interactions may be separated on the
basis of overall event characteristics.  The signatures that can be
used to distinguish between production processes are whether there are
rapidity gaps in the event and whether the nuclei break up. Nuclear
breakup can be measured with downstream ZDCs.

Other variables may be helpful in separating event classes.  The
event multiplicity is lower for photoproduction and two-photon
interactions because less energy is available.  The
multiplicity also depends on the details of the interaction.  For
two-photon interactions, the total event $p_T$ should be less than
$2\hbar c/R_A$.  Unfortunately, it is necessary to reconstruct the
entire event to measure the total $p_T$.  This is
difficult in $Q\overline Q$ events.  Because of these difficulties,
multiplicity and $p_T$ will not be further considered here as a
separation factor.

This section will focus on isolating clean photoproduction and
two-photon final states.  The large hadroproduction cross sections are
a background to these events.  The cuts discussed here are geared
toward reducing the hadroproduction background and thereby
differentiating between the three production processes.  RHIC
data\cite{ZDCanalysis} show that almost all hadronic interactions
break up both nuclei.  In contrast, photoproduction should only
dissociate the target nucleus while two-photon interactions should
leave both nuclei intact.  However, the photoproduction and two-photon
interactions occur at moderate impact parameters, less than 50 (500)
fm at RHIC (LHC), where one or more additional photons may be
exchanged.

For heavy nuclei like lead or gold, the additional photons can excite
one or both nuclei, leading to nuclear breakup.  Except for the common
impact parameter, these additional photons are independent of the
two-photon or photonuclear events\cite{bkn}.  The probabilities for
excitation are significant.  The probability of a single given nucleus
being excited in a collision at $b=2R_A$ is about 30\% with gold at
RHIC, rising to 35\% for lead at the LHC\cite{baurrev}.  As $b$ rises,
the excitation probability drops as $1/b^2$.  The breakup probability
scales as $A^{10/3}$ so that for even slightly lighter nuclei like
argon or silicon the breakup probability is considerably reduced.
Since the nuclear breakup probabilities are independent of each other,
the probability for both nuclei to dissociate is simply the square of
the single dissociation probability.

One way to select photoproduction events is to eliminate events where
both nuclei break up.  This should
eliminate almost all of the hadroproduction events while retaining
most of the photoproduction and two-photon interactions.
Unfortunately, there are no calculations of the hadronic interaction
cross sections without accompanying nuclear breakup. Indeed, such a
calculation is problematic because too little is known about the
recoil energy transfer in the nucleus.  However, using a Glauber
calculation, we find that the cross section for a single
nucleon-single nucleon interaction in an Au+Au collision is about 700
mb, 10\% of the total hadronic cross section.  At RHIC and LHC
energies even soft nuclear interactions involve significant energy
transfer.  Thus phase space considerations dictate that the
interacting nucleons are likely to be ejected from the nucleus.  There
could then be considerable momentum transferred to the nuclear
fragments.   Here we assume that each nucleus has a 10\% chance of remaining
intact.  With these assumptions, the heavy quark hadroproduction cross
sections with one nucleus remaining intact are not too different from
those presented in Section \ref{hadrosec}.  The 10\% chance of
remaining intact may be high, but, to be conservative, we will use the
rates in Section \ref{hadrosec} to calculate the hadronic backgrounds
to photoproduction.

The rejection factor $R$ is the probability of finding a rapidity gap 
with width $y$ in a hadronic collision, $R \sim \exp(-y dN/dy)$, where
$dN/dy$ is the average multiplicity per unit rapidity. Here, we will
only consider charged particles but with a calorimeter to detect
neutrals, the rejection would be enhanced.  The UA1 collaboration
parameterized the charged particle multiplicity as $dN_{\rm
ch}/dy\approx -0.32+\ln{\sqrt{S/{\rm GeV}}}$\cite{ua1}.  Neglecting
the small difference between $p\overline p$ and $pp$ collisions, at
midrapidity at RHIC, $dN_{\rm ch}/dy\approx 2.6$ \cite{stara}, rising
to $dN_{\rm ch}/dy \approx 4.4$ at the LHC.  Most modern experiments
use forward detectors like beam-beam counters to measure particle
production over a wide rapidity range. These counters can be used to
find rapidity gaps.  Here we will consider two charged particle
detectors each covering 2 units of rapidity on each side of a central
detector, representative of modern experiments.  We scale the
midrapidity multiplicities by 0.6 because $dN_{\rm ch}/dy$
decreases away from midrapidity.

For photoproduction, requiring a single rapidity gap will reject all
but $R = \exp{(-2 \times 0.6 \times 2.6)} = 0.04$ of the hadronic
interactions at RHIC while at LHC the rejection for a similar
detector is $R=\exp{(-2 \times 0.6 \times 4.4)} = 0.005$. These
factors, calculated for $pp$ collisions, should be conservative for
$pA$ and $AA$ collisions where there is typically more than a single
nucleon-single nucleon collision.  These factors would also apply to
the rejection of photoproduction backgrounds when considering
two-photon reactions.  Since there are two rapidity gaps in two-photon
interactions, these
rejection factors are squared.

The CDF collaboration has experimentally observed comparable rejection
factors in a study of diffractive bottom production\cite{CDF}.  They
isolated a diffractively produced bottom sample from $p\overline p$
collisions\cite{CDF} with a signal to noise ratio of 3:1 despite the
fact that diffractive production was 1/160 of the total hadronic $b$
cross section.  This corresponded to $R=0.002$, smaller than the $R$ values
calculated above.  This is probably because the CDF forward counters
cover 2.7 units of rapidity and are supplemented with forward
calorimeters that detect neutrals.  Thus the rejection factors
are not unrealistic and could even be improved on with more solid angle
coverage.  Of course, for nuclear beams the higher multiplicities per
participant pair should increase the rejection factors, even for
single nucleon-single nucleon interactions, due to possible secondary
interactions.

The rapidity gap requirements should reject relatively few signal
events since photoproduction always leads to a rapidity gap.  The
exceptions are the events that are accompanied by additional
electromagnetic interactions.  A small fraction of these breakups will
involve high energy photons which produce final state particles that
fill in the rapidity gap, causing the signal event to be lost.  Even
at the LHC, the breakup probability due to high energy photons is
quite small and events with additional particles should not
appreciably affect the signal.  If it is necessary to also require
that one nucleus remains intact, then signal loss will need to be
considered but such loss will not affect the viability of the
measurement.

Other backgrounds are neglected here.  Examples include single
diffractive (hadronic) charm production and double-Pomeron
interactions.  Single diffractive production will have one rapidity
gap and could potentially mimic charm photoproduction.  However, the
diffractive final state would be produced quite near the beam
rapidity, forward of the predominantly photoproduced charm.
Double-Pomeron charm production will be in the central region with
two rapidity gaps.  Because the Pomeron has such a short range, both
single and double diffractive interactions can only occur over a very
narrow range of impact parameters so that their $AA$ cross sections
should be small \cite{pompom}.  The other major background, beam-gas
interactions, is detector and vacuum system specific and will not be
discussed here.

Conservatively, both the rapidity gaps and nuclear breakup criteria
should reject more than $99\%$ of the hadronic events.  Although these
criteria are not completely independent, comparing the numbers in
Sections \ref{photosec} - 
\ref{gamgamsec} shows that application of either criteria should lead
to a good signal to noise ratio for selecting either photoproduction
or two-photon interactions.  If 99\% of the hadronic events with
$b>2R_A$ can be rejected, then hadroproduction is only a small
background to photoproduction of heavy quarks, one that can
be controlled to the degree necessary to measure shadowing by
comparing $pA$ and $AA$.

One could also select events with two rapidity gaps and no nuclear
breakup to search for two-photon interactions. However, in almost all
cases, the two-photon cross sections are a factor of at least 1,000
smaller than the photoproduction cross sections.  This factor is
smaller than the single-gap $R$ calculated above as well as larger than
rejection obtained by CDF and unlikely to be achieved in a real
experiment.  Selecting two-photon events may require completely
reconstructing the events and using the low total $p_T$ of the event
as a final selection criteria.  This reconstruction will necessarily 
have a very low overall efficiency, thus requiring very large data sets.

In conclusion, both rapidity gaps and the absence of nuclear breakup
are effective criteria to separate photoproduction interactions.
Charged-particle multiplicity and other event characteristics may also
be useful in refining the event selection.  Once a sample of
photoproduction events have been isolated, charm and bottom production
may be studied with conventional selection techniques such as prompt leptons,
separated vertices, and $D$ or $B$ meson reconstruction.

\section{Conclusions}
\label{sumsec}

We have calculated the total cross sections, quark $p_T$ and $y$
distributions, and pair mass spectra for hadronic, photonuclear and
two-photon production of heavy quark pairs using a consistent set of
quark masses and scales.  The hadronic processes have the largest
cross sections, followed by photoproduction and two-photon
interactions.  However, using the characteristics of rapidity gaps
and nuclear breakup, photoproduction and two-photon processes
should be cleanly separable.

Photoproduction and two-photon measurements can be used to test the
QCD calculations.  Shadowing has a 10\% effect on $c\overline c$
production with heavy nuclei at RHIC, rising to 20\% at the LHC.  By
comparing the production cross sections in $pA$ and $AA$ collisions,
most theoretical uncertainties cancel so that shadowing can be
accurately measured if the photon flux is well known.  By using
appropriate calibration signals, it appears that the photon flux
uncertainties can be understood and useful gluon shadowing
measurements made.

R.V. thanks the Gesellschaft f\" ur Schwerionenforschung, the Niels
Bohr Institute, and the Grand Acc\' el\' erateur National d'Ions
Lourds for hospitality while this work was in progress.  This work was
supported in part by the Division of Nuclear Physics of the Office of
High Energy and Nuclear Physics of the U.S. Department of Energy under
Contract No. DE-AC-03-76SF00098 and by the Swedish Research Council
(VR).

\begin{table}
\begin{tabular}{ccccc}
& \multicolumn{2}{c}{$AA$} & \multicolumn{2}{c}{$pA$} \\
$A$       & $\sqrt{S_{NN}}$ (GeV) & ${\cal L}_{AA}$ (cm$^{-2}$s$^{-1}$) &
$\sqrt{S_{NN}}$ (GeV) & ${\cal L}_{pA}$ (cm$^{-2}$s$^{-1}$) \\ \hline
\multicolumn{5}{c}{RHIC} \\
O       & 250 & $9.8\times10^{28}$  & 250  & $1.2\times10^{30}$ \\
Si      & 250 & $4.4\times10^{28}$  & 250  & $8\times10^{29}$ \\
I       & 208 & $2.7\times10^{27}$  & 208  & $2\times10^{29}$ \\
Au      & 200 & $2\times10^{26}$    & 200  & $6\times10^{28}$ \\
\multicolumn{5}{c}{LHC} \\
O       & 7000 & $1.6\times10^{29}$ & 9900 & $1.0\times10^{31}$ \\
Ar      & 6300 & $4.3\times10^{28}$ & 9390 & $5.8\times10^{30}$ \\
Pb      & 5500 & $4.2\times10^{26}$ & 8800 & $7.4\times10^{29}$ \\
\end{tabular}
\caption[]{Luminosities and beam energies for $AA$ and $pA$ collisions
at RHIC and LHC.}
\label{lum}
\end{table}

\begin{table}
\begin{tabular}{cccccc}
\multicolumn{6}{c}{$AA$} \\ 
$A$       & $E_{\rm beam}$ (GeV) & $\gamma_L$ & $k_{\rm max}$ (GeV) &
$E_{\rm max}$ (TeV) & $\sqrt{S_{\rm max}}$ (GeV) \\ \hline
\multicolumn{6}{c}{RHIC} \\
O       & 125 & 133  & 12.7  & 2.31  & 66 \\
Si      & 125 & 133  &  8.5  & 1.92  & 60 \\
I       & 104 & 111  &  3.9  & 0.81  & 39 \\
Au      & 100 & 106  &  3.2  & 0.66  & 35 \\
\multicolumn{6}{c}{LHC} \\
O       & 3500 & 3730 & 357 & 1820 & 1850 \\
Ar      & 3150 & 3360 & 185 & 1080 & 1430 \\
Pb      & 2750 & 2930 &  87.0 &  480 &  950 \\ \hline
\multicolumn{6}{c}{$pA$ LHC} \\
O       & 4950 & 5270 & 435 & 3630 & 2610 \\
Ar      & 4700 & 5000 & 276 & 2410 & 2130 \\
Pb      & 4400 & 4690 & 139 & 1220 & 1500 \\
\end{tabular}
\caption[]{Beam energies, $E_{\rm beam}$, Lorentz factors, $\gamma_L$,
photon cutoff energy in the center of mass frame, $k_{\rm max}$, and
in the nuclear rest frame, $E_{\rm max}$, and equivalent
nucleon-nucleon center of mass energy, $\sqrt{S_{\rm max}}$, for $AA$
collisions at RHIC and the LHC.  Since the ion and proton beam
energies are expected to be the same in $pA$ and $AA$ collisions at
RHIC, we only distinguish the $pA$ energies at LHC.}
\label{gamfacs}
\end{table}

\begin{table}
\begin{tabular}{ccccc}
$AA$ & $\sigma^{\rm dir}(S^i=1)$ (mb) & $\sigma^{\rm dir}({\rm EKS98})$ (mb) 
& $\sigma^{\rm res}(S^i=1)$ (mb)
& $\sigma^{\rm res}({\rm EKS98})$ (mb) \\ \hline
\multicolumn{5}{c}{RHIC} \\
O+O   & 0.067  & 0.068  & 0.0019  & 0.0019 \\
Si+Si & 0.30   & 0.31   & 0.0080  & 0.0083 \\
I+I   & 8.96   & 9.74   & 0.199   & 0.206  \\
Au+Au & 24.8  & 27.4  & 0.530   & 0.550  \\
\multicolumn{5}{c}{LHC} \\
O+O   & 2.35   & 2.13   & 0.351   & 0.346  \\
Ar+Ar & 23.3  & 20.4  & 3.00    & 2.93   \\
Pb+Pb & 1790 & 1500 & 190.0   & 186.7  \\
\end{tabular}
\caption[]{Direct and resolved $c\overline c$ photoproduction
cross sections integrated over $b>2R_A$ in peripheral $AA$ collisions
at RHIC and LHC.  }
\label{gamAcc}
\end{table}

\begin{table}
\begin{tabular}{ccccc}
$AA$ & $\sigma^{\rm dir}(S^i=1)$ ($\mu$b) & $\sigma^{\rm dir}({\rm EKS98})$ 
($\mu$b) & $\sigma^{\rm res}(S^i=1)$ ($\mu$b)
& $\sigma^{\rm res}({\rm EKS98})$ ($\mu$b) \\ \hline
\multicolumn{5}{c}{RHIC} \\
O+O   & 0.047 & 0.049 & 0.0031 & 0.0031 \\
Si+Si & 0.178 & 0.188 & 0.0116 & 0.0115 \\
I+I   & 2.33  & 2.46  & 0.154  & 0.148  \\
Au+Au & 4.94  & 5.22  & 0.332  & 0.315  \\
\multicolumn{5}{c}{LHC} \\
O+O   & 11.9 & 11.4 & 2.93   & 2.93 \\
Ar+Ar & 107  & 102  & 22.2   & 22.6 \\
Pb+Pb & 718  & 686  & 121    & 126 \\
\end{tabular}
\caption[]{Direct and resolved $b \overline b$ photoproduction
cross sections integrated over $b > 2R_A$ in peripheral $AA$ collisions
at RHIC and LHC.}
\label{gamAbb}
\end{table}

\begin{table}
\begin{tabular}{ccccc}
& \multicolumn{2}{c}{$c \overline c$} & \multicolumn{2}{c}{$b \overline b$} \\
$pA$ & $\sigma^{\rm dir}(S^i=1)$ ($\mu$b) & $\sigma^{\rm res}(S^i=1)$ ($\mu$b) 
& $\sigma^{\rm dir}(S^i=1)$ (nb) & $\sigma^{\rm res}(S^i=1)$ (nb) \\ \hline
\multicolumn{5}{c}{RHIC} \\ 
$p$O  & 2.68  & 0.081 & 2.34 & 0.154 \\
$p$Si & 7.29  & 0.213 & 5.79 & 0.378 \\
$p$I  & 54.2  & 1.32  & 23.5 & 1.52  \\
$p$Au & 100   & 2.34  & 35.5 & 2.31  \\
\multicolumn{5}{c}{LHC} \\
$p$O  & 110  & 20.7 & 630 & 202 \\
$p$Ar & 485 & 83.9 & 2670  & 774 \\
$p$Pb & 7940  & 1190  & 40100 & 9910  \\
\end{tabular}
\caption[]{Direct and resolved $c\overline c$ and $b \overline b$ 
photoproduction cross sections integrated over $b > r_p + R_A$ 
in $pA$ collisions at RHIC and LHC.  }
\label{gampccbb}
\end{table}

\begin{table}
\begin{tabular}{ccccc}
& \multicolumn{2}{c}{$c \overline c$} & \multicolumn{2}{c}{$b \overline b$} \\
$AA$ & $N(S^i=1)$ & $N({\rm EKS98})$ & $N(S^i=1)$ & $N({\rm EKS98})$ \\ \hline
\multicolumn{5}{c}{RHIC} \\ 
O+O   & $6.75\times 10^6$ & $6.94\times 10^6$ & $4.88\times 10^3$ 
& $5.10\times 10^3$ \\
Si+Si & $1.36\times 10^7$ & $1.41\times 10^7$ & $8.35\times 10^3$ 
& $8.73\times 10^3$ \\
I+I   & $2.47\times 10^7$ & $2.69\times 10^7$ & $6.70\times 10^3$ 
& $7.06\times 10^3$ \\
Au+Au & $5.07\times 10^6$ & $5.60\times 10^6$ & $1.06\times 10^3$ 
& $1.10\times 10^3$ \\
\multicolumn{5}{c}{LHC} \\
O+O   & $4.15\times 10^8$ & $3.80\times 10^8$ & $2.29\times 10^6$ 
& $2.20\times 10^6$ \\
Ar+Ar & $1.13\times 10^9$ & $9.98\times 10^8$ & $5.58\times 10^6$ 
& $5.37\times 10^6$ \\
Pb+Pb & $8.29\times 10^8$ & $7.05\times 10^8$ & $3.58\times 10^5$ 
& $3.46\times 10^5$ \\
\end{tabular}
\caption[]{Total $c\overline c$ and $b \overline b$ photoproduction
rates in peripheral $AA$ collisions over a $10^6$ s run at RHIC and LHC.  
The rates are based on Tables~{\protect \ref{gamAcc}} and 
{\protect \ref{gamAbb}}.}
\label{NgamAccbb}
\end{table}

\begin{table}
\begin{tabular}{ccc}
& $c \overline c$ & $b \overline b$ \\
$pA$ & $N(S^i=1)$ & $N(S^i=1)$ \\ \hline
\multicolumn{3}{c}{RHIC} \\
$p$O  & $3.30\times 10^6$ & $2.99\times 10^3$ \\
$p$Si & $6.00\times 10^6$ & $4.93\times 10^3$ \\
$p$I  & $1.11\times 10^5$ & $5.00\times 10^3$ \\
$p$Au & $6.08\times 10^5$ & $2.27\times 10^3$ \\
\multicolumn{3}{c}{LHC} \\
$p$O  & $1.32\times 10^9$ & $8.32\times 10^6$ \\
$p$Ar & $3.11\times 10^9$ & $1.88\times 10^7$ \\
$p$Pb & $6.70\times 10^9$ & $3.69\times 10^7$ \\
\end{tabular}
\caption[]{Total $c\overline c$ and $b \overline b$ photoproduction
rates in $pA$ collisions over a $10^6$ s run at RHIC and LHC.  The rates 
are based on Table~{\protect \ref{gampccbb}}.}
\label{Ngampccbb}
\end{table}

\begin{table}
\begin{tabular}{cccc}
$AA$ & $\sigma(S^i=1)$ (mb) & $\sigma({\rm EKS98})$ 
(mb) & $\sigma({\rm EKS98}b)$ (mb) \\ \hline
\multicolumn{4}{c}{RHIC} \\
O+O   & 4.04 & 3.93 & 4.00 \\
Si+Si & 8.54 & 8.22 & 8.42 \\
I+I   & 22.6 & 21.6 & 22.3 \\
Au+Au & 33.1 & 31.6 & 32.6  \\
\multicolumn{4}{c}{LHC} \\
O+O   & 113  & 93.2  & 104 \\
Ar+Ar & 426  & 323   & 379 \\
Pb+Pb & 1090 & 714   & 948 \\
\end{tabular}
\caption[]{Total $c\overline c$ hadroproduction cross sections integrated over
$b > 2R_A$ in
peripheral $AA$ collisions at RHIC and LHC.  }
\label{AAccbar}
\end{table}

\begin{table}
\begin{tabular}{cccc}
$AA$ & $N(S^i=1)$ & $N({\rm EKS98})$ & $N({\rm EKS98}b)$ \\ \hline
\multicolumn{4}{c}{RHIC} \\
O+O   & $3.96\times 10^8$ & $3.85\times 10^8$ & $3.92\times 10^8$ \\
Si+Si & $3.76\times 10^8$ & $3.62\times 10^8$ & $3.70\times 10^8$ \\
I+I   & $6.10\times 10^7$ & $5.84\times 10^7$ & $6.02\times 10^7$ \\
Au+Au & $6.62\times 10^6$ & $6.33\times 10^6$ & $6.52\times 10^6$  \\
\multicolumn{4}{c}{LHC} \\
O+O   & $1.74\times 10^{10}$ & $1.44\times 10^{10}$ & $1.61\times 10^{10}$ \\
Ar+Ar & $1.83\times 10^{10}$ & $1.39\times 10^{10}$ & $1.63\times 10^{10}$ \\
Pb+Pb & $4.57\times 10^8$    & $3.00\times 10^8$    & $3.98\times 10^8$    \\
\end{tabular}
\caption[]{Total $c\overline c$ hadroproduction rates in peripheral
$AA$ collisions at RHIC and LHC with a $10^6$ s run.  The rates are based on 
Table~{\protect \ref{AAccbar}}.}
\label{NAAccbar}
\end{table}

\begin{table}
\begin{tabular}{cccc}
$AA$ & $\sigma(S^i=1)$ ($\mu$b) & $\sigma({\rm EKS98})$ ($\mu$b) & 
$\sigma({\rm EKS98}b)$ ($\mu$b) \\ \hline
\multicolumn{4}{c}{RHIC} \\
O+O   & 22.7 & 24.1 & 23.3 \\
Si+Si & 47.9 & 51.7 & 49.5 \\
I+I   & 111 & 127 & 117 \\
Au+Au & 154 & 183 & 167  \\
\multicolumn{4}{c}{LHC} \\
O+O   & 2490  & 2260  & 2390 \\
Ar+Ar & 9110  & 7930  & 8600 \\
Pb+Pb & 21700 & 17500 & 20200 \\
\end{tabular}
\caption[]{Total $b\overline b$ hadroproduction cross sections integrated over
$b > 2R_A$ in
peripheral $AA$ collisions.  }
\label{AAbbbar}
\end{table}

\begin{table}
\begin{tabular}{cccc}
$AA$ & $N(S^i=1)$ & $N({\rm EKS98})$ & $N({\rm EKS98}b)$ \\ \hline
\multicolumn{4}{c}{RHIC} \\
O+O   & $2.22\times 10^6$ & $2.36\times 10^6$ & $2.27\times 10^6$ \\
Si+Si & $2.11\times 10^6$ & $2.27\times 10^6$ & $2.18\times 10^6$ \\
I+I   & $3.01\times 10^5$ & $3.43\times 10^5$ & $3.16\times 10^5$ \\
Au+Au & $3.17\times 10^4$ & $3.67\times 10^4$ & $3.34\times 10^4$  \\
\multicolumn{4}{c}{LHC} \\
O+O   & $3.84\times 10^8$ & $3.48\times 10^8$ & $3.68\times 10^8$ \\
Ar+Ar & $3.92\times 10^8$ & $3.41\times 10^8$ & $3.70\times 10^8$ \\
Pb+Pb & $9.11\times 10^6$ & $7.35\times 10^6$ & $8.48\times 10^6$    \\
\end{tabular}
\caption[]{Total $b\overline b$ hadroproduction
rates in peripheral $AA$ collisions with a $10^6$ s run
at RHIC and LHC.  The rates are based on Table~{\protect \ref{AAbbbar}}.}
\label{NAAbbbar}
\end{table}

\begin{table}
\begin{tabular}{ccccc}
& \multicolumn{2}{c}{$c \overline c$} & \multicolumn{2}{c}{$b \overline b$} \\
$pA$ & $\sigma(S^i=1)$ (mb) & $\sigma({\rm EKS98})$ (mb) & $\sigma(S^i=1)$ ($\mu$b)
& $\sigma({\rm EKS98})$ ($\mu$b) \\ \hline
\multicolumn{5}{c}{RHIC} \\
$p$O   & 4.19  & 4.14  & 23.7 & 24.4 \\
$p$Si  & 7.33  & 7.21  & 41.5 & 43.1 \\
$p$I   & 26.2 & 25.8 & 131 & 139 \\
$p$Au  & 38.6 & 38.1 & 187 & 201 \\
\multicolumn{5}{c}{LHC} \\
$p$O   & 153  & 138  & 3740  & 3540 \\
$p$Ar  & 368  & 318  & 8850  & 8180 \\
$p$Pb  & 1820 & 1460 & 43000 & 38200 \\
\end{tabular}
\caption[]{Total $c\overline c$ and $b \overline b$ hadroproduction
cross sections in minimum bias (all $b$) $pA$ collisions
at RHIC and LHC. }
\label{pAccbb}
\end{table}

\begin{table}
\begin{tabular}{ccccc}
& \multicolumn{2}{c}{$c \overline c$} & \multicolumn{2}{c}{$b \overline b$} \\
$pA$ & $N(S^i=1)$ & $N({\rm EKS98})$ & $N(S^i=1)$ & $N({\rm EKS98})$ \\ \hline
\multicolumn{5}{c}{RHIC} \\
$p$O   & $5.02\times 10^9$ & $4.96\times 10^9$ & $2.85\times 10^7$ 
& $2.92\times 10^7$ \\
$p$Si  & $5.86\times 10^9$ & $5.77\times 10^9$ & $3.32\times 10^7$ 
& $3.45\times 10^7$ \\
$p$I   & $5.25\times 10^9$ & $5.17\times 10^9$ & $2.62\times 10^7$ 
& $2.79\times 10^7$ \\
$p$Au  & $2.32\times 10^9$ & $2.28\times 10^9$ & $1.12\times 10^7$ 
& $1.20\times 10^7$ \\
\multicolumn{5}{c}{LHC} \\
$p$O   & $1.53\times 10^{12}$ & $1.39\times 10^{12}$ & $3.75\times 10^{10}$ 
& $3.57\times 10{10}$ \\
$p$Ar  & $2.01\times 10^{12}$ & $1.75\times 10^{12}$ & $4.87\times 10^{10}$ 
& $4.50\times 10^{10}$ \\
$p$Pb  & $1.35\times 10^{12}$ & $1.07\times 10^{12}$ & $3.18\times 10^{10}$ 
& $2.82\times 10^{10}$ \\
\end{tabular}
\caption[]{Total $c\overline c$ and $b \overline b$ hadroproduction
rates in minimum bias $pA$ collisions over a $10^6$ s run at RHIC and LHC.  
The rates are based on Table~{\protect \ref{pAccbb}}.}
\label{NpAccbb}
\end{table}

\begin{table}
\begin{tabular}{ccccccc}
& \multicolumn{3}{c}{all $c \overline c$} & \multicolumn{3}{c}{$M > 2m_D$} \\
$AA$ & $\sigma^{\rm dir}$ (nb) & $\sigma^{\rm 1-res}$ (nb) 
& $\sigma^{\rm 2-res}$ (nb) &
$\sigma^{\rm dir}$ (nb) & $\sigma^{\rm 1-res}$ (nb) 
& $\sigma^{\rm 2-res}$ (nb)\\ \hline
\multicolumn{7}{c}{RHIC} \\
O+O   & 4.64    & 0.08   & 0.0020 & 1.65  &  0.022  & 0.00039 \\
Si+Si & 32.0    & 0.49  & 0.013  & 10.8  &  0.125  & 0.0023  \\
I+I   & 1320    & 10.7  & 0.345  & 288 &  1.18   & 0.027   \\
Au+Au & 3650    & 22.2  & 0.786  & 601 &  1.37   & 0.035   \\
\multicolumn{7}{c}{LHC} \\
O+O   & 236     & 11.7   & 0.24   & 128   &  6.01   & 0.10    \\
Ar+Ar & 4530    & 210.0  & 4.36   & 2410  &  105    & 1.76    \\
Pb+Pb & 1110000 & 45000  & 951    & 565000&  21400  & 352   \\
\end{tabular}
\caption[]{Two photon $c\overline c$ cross sections in peripheral $AA$
collisions at RHIC and LHC, integrated over $b>2R_A$.}
\label{gamgamcc}
\end{table}

\begin{table}
\begin{tabular}{ccccccc}
& \multicolumn{3}{c}{all $b \overline b$} & \multicolumn{3}{c}{$M > 2m_B$} \\
$AA$ & $\sigma^{\rm dir}$ (pb) & $\sigma^{\rm 1-res}$ (pb) 
& $\sigma^{\rm 2-res}$ (pb) &
$\sigma^{\rm dir}$ (pb) & $\sigma^{\rm 1-res}$ (pb) 
& $\sigma^{\rm 2-res}$ (pb)\\ \hline
\multicolumn{7}{c}{RHIC} \\
O+O   & 0.268 & 0.0018 & 0.00038 & 0.194 & 0.0010 & 0.00029 \\
Si+Si & 0.923 & 0.0031 & 0.00083 & 0.582 & 0.0013 & 0.00046 \\
\multicolumn{7}{c}{LHC} \\
O+O   & 285  & 31.7  & 3.08  & 262.6  & 28.9  & 2.62  \\
Ar+Ar & 4890   & 491.0 & 49.3  & 4480   & 444 & 41.7  \\
Pb+Pb & 943000 & 75000 & 8260  & 855000 & 66800 & 6820  \\
\end{tabular}
\caption[]{Two photon $b\overline b$ cross sections in peripheral $AA$
collisions at RHIC and LHC, integrated over $b>2R_A$.}
\label{gamgambb}
\end{table}

\begin{table}
\begin{tabular}{ccccc}
& \multicolumn{2}{c}{all $Q \overline Q$} & \multicolumn{2}{c}{$M > 2m_H$} \\
$AA$ & $N(c \overline c)$ & $N(b \overline b)$ & $N(c
\overline c)$ & $N(b \overline b)$ \\ \hline
\multicolumn{5}{c}{RHIC} \\
O+O   & $4.62\times 10^2$ & $2.65\times 10^{-2}$ & $1.64\times 10^2$ 
& $1.91\times 10^{-2}$ \\
Si+Si & $1.43\times 10^3$ & $4.08\times 10^{-2}$ & $4.81\times 10^2$ 
& $2.57\times 10^{-2}$ \\
I+I   & $3.61\times 10^3$ & - & $7.81\times 10^2$ & -  \\
Au+Au & $7.36\times 10^2$ & - & $1.21\times 10^2$ & -   \\
\multicolumn{5}{c}{LHC} \\
O+O   & $3.97\times 10^4$ & $5.11\times 10^1$ & $2.15\times 10^4$ 
& $4.71\times 10^1$\\
Ar+Ar & $2.04\times 10^5$ & $2.33\times 10^2$ & $1.08\times 10^5$ 
& $2.13\times 10^2$ \\
Pb+Pb & $4.84\times 10^5$ & $4.41\times 10^2$ & $2.47\times 10^5$ 
& $3.90\times 10^2$ \\
\end{tabular}
\caption[]{Total $c\overline c$ and $b \overline b$ two-photon rates
in peripheral $AA$ collisions over a $10^6$ s run at RHIC and LHC.  
The rates are based on Tables~{\protect \ref{gamgamcc}} and 
{\protect \ref{gamgambb}}.}
\label{Ngamgamccbb}
\end{table}

\newpage

\newpage

\newpage

\begin{references}
\def\etal{{\it et al.}}

\bibitem{baurrev}G. Baur, K. Hencken and D. Trautmann, J. Phys. G {\bf 24},
1457 (1998); G. Baur, K. Hencken, D. Trautmann, S. Sadovsky and Y. Kharlov,
Phys. Rep. {\bf 364}, 359 (2002).

\bibitem{ZDCanalysis}M. Chiu {\it et al.}, Phys. Rev. Lett. {\bf 89},
012302 (2002).

\bibitem{usPRC}S.R. Klein and J. Nystrand, Phys. Rev. C {\bf 60}, 014903
(1999).

\bibitem{parkcity}C. Adler {\it et al.} (STAR Collaboration),
nucl-ex/0206004, submitted to Phys. Rev. Lett; S. Klein (STAR
Collab.), in proceedings of the {\it 17$^{\rm th}$ Winter Workshop on
Nuclear Dynamics}, Park City, UT, USA, 2001, nucl-ex/0104016.

\bibitem{INPC}S. Klein (STAR Collab.), in proceedings of {\it INPC 2001},
Berkeley, CA, USA, 2001, nucl-ex/0100018.

\bibitem{FELIX}E. Lippmaa \etal, {\it FELIX: A full acceptance
detector at the LHC}, CERN/LHCC 97-45, August, 1997.

\bibitem{kp} K. Piotrzkowski, Phys. Rev. D {\bf 63}, 071502 (2001).

\bibitem{bbar}N. Baron and G. Baur, Phys. Rev. C {\bf 48}, 1999
(1993).

\bibitem{bbar2}M. Greiner, M. Vidovi\'c, C. Hofmann, A. Sch\"afer
and G. Soff, Phys. Rev. C {\bf 51}, 911 (1995).

\bibitem{knv1} S.R. Klein, J. Nystrand and R. Vogt, Eur. Phys.
J. C {\bf 21}, 563 (2001).

\bibitem{vidovictwophoton}M. Vidovi\'c, M. Greiner and G. Soff,
J. Phys. G {\bf 21}, 545 (1995).

\bibitem{hpc} P.L. McGaughey {\it et al.}, Int. J. Mod. Phys. A {\bf 10},
2999 (1995). 

\bibitem{frixione}S. Frixione, P. Nason and G. Ridolfi, Nucl. Phys.
B {\bf 454}, 3 (1995).

\bibitem{berezhnoy}A.V. Berezhnoy, V.V. Kiselev and A.K. Likhoded,
Phys. Rev. D {\bf 62}, 074013 (2000). 

\bibitem{Drees} M. Drees, M. Kramer, J. Zunft and P.M. Zerwas, 
Phys. Lett. B {\bf 306}, 371 (1993).

\bibitem{l3}M. Acciarri {\it et al.} (L3 Collab.),
Phys. Lett. B {\bf 453}, 83 (1999).

\bibitem{CDF}T. Affolder {\it et al.} (CDF Collab.), Phys. Rev. Lett.
{\bf 84}, 232 (2000).

\bibitem{harris} E.L. Berger, B.W. Harris, D.E. Kaplan, Z. Sullivan, T.M.P. 
Tait and C.E.M. Wagner, Phys. Rev. Lett. {\bf 86}, 4231 (2001).

\bibitem{other} A.P. Lipatov, V.A. Saleev and N.P. Zotov, hep-ph/0112114.

\bibitem{PDG} D.E. Groom {\it et al.} (Particle Data Group), Eur. Phys. J. C
{\bf 15}, 1 (2000).

\bibitem{RHIClum} RHIC Conceptual Design Report (BNL-52195) 1989.

\bibitem{brandt}D. Brandt, LHC Project Report 450, December, 2000.

\bibitem{brandt2} D. Brandt, in Proc. of the $6^{\rm th}$ CMS Heavy Ion
Workshop, MIT, February 2002.

\bibitem{bertulani}V.P. Goncalves and C.A. Bertulani,
Phys. Rev. C {\bf 65}, 054905 (2002).

\bibitem{witten} E. Witten, Nucl. Phys. B {\bf 120}, 189 (1977).

\bibitem{jpgconf} T. Sj\"ostrand, J.K. Storrow, and A. Vogt, J. Phys. G {\bf 
22}, 893 (1996).

\bibitem{joneswyld} L.M. Jones and H.W. Wyld, Phys. Rev. D {\bf 17}, 759 
(1978).

\bibitem{usinterf}S.R. Klein and J. Nystrand, Phys. Rev. Lett.
{\bf 84}, 2330 (2000).

\bibitem{SvN} J. Smith and W.L. van Neerven, Nucl. Phys. B {\bf 374}, 
36 (1992).

\bibitem{GRVgam} M. Gl\"{u}ck, E. Reya, and A. Vogt, Phys. Rev. D {\bf 46}, 
1973 (1992); Phys. Rev. D {\bf 45}, 3986 (1992).

\bibitem{DG1} M. Drees and K. Grassie, Z. Phys. C {\bf 28}, 451 (1985).

\bibitem{LAC1} H. Abramowicz, K. Charchula, and A. Levy, Phys. Lett. B {\bf 
269}, 458 (1991).

\bibitem{WHIT} K. Hagiwara, M. Tanaka, I. Watanabe, and T. Izubuchi, Phys. Rev.
D {\bf 51}, 3197 (1995).

\bibitem{SaS} G.A. Schuler and T. Sj\"ostrand, Z. Phys. C {\bf 68}, 607 (1995);
Phys. Lett. B {\bf 276}, 193 (1996).

\bibitem{PDFLIB} H. Plothow-Besch,  `PDFLIB: Proton, Pion
and Photon Parton Density Functions, Parton Density Functions of the Nucleus, 
and $\alpha_s$ Calculations',
User's Manual - Version 8.04, W5051 PDFLIB, 2000.04.17, CERN-ETT/TT.

\bibitem{JADE} W. Bartel {\em et al.} (JADE Collab.), Z. Phys. C {\bf 24}, 231
(1984).

\bibitem{CTEQRMP} G. Sterman {\em et al.} (CTEQ Collab.), 
Rev. Mod. Phys. {\bf 67}, 157 (1995).

\bibitem{spenprc} V. Emel'yanov, A. Khodinov, S.R. Klein and R. Vogt,
Phys. Rev. C {\bf 56}, 2726 (1997).

\bibitem{spenprl} V. Emel'yanov, A. Khodinov,
S.R. Klein and R. Vogt, Phys. Rev. Lett {\bf 81}, 1801 (1998).

\bibitem{spenpsi} V. Emel'yanov, A. Khodinov, S.R. Klein, and 
R. Vogt, Phys. Rev. C {\bf 59}, 1860 (1999).

\bibitem{ekkv4} V. Emel'yanov, A. Khodinov, S.R. Klein and R. Vogt,
Phys. Rev. C {\bf 61}, 044904 (2000).

\bibitem{RVwz} R. Vogt, Phys. Rev. C {\bf 64}, 044901 (2001).

\bibitem{pdf}A.D. Martin, R.G. Roberts, W.J. Stirling and R.S.
Thorne, Eur. Phys. J. C {\bf 4}, 463 (1998); Phys. Lett. B {\bf 443} 301
(1998).

\bibitem{Vvv}
C.W. deJager, H. deVries and C. deVries, Atomic Data and Nuclear Data
Tables {\bf 14}, 485 (1974).

\bibitem{eskola} K.J. Eskola, V.J. Kolhinen and P.V. Ruuskanen,
Nucl. Phys. B {\bf 535}, 351 (1998); K.J. Eskola, V.J. Kolhinen and
P.V. Ruuskanen, Eur. Phys. J. C {\bf 9}, 61 (1999).

\bibitem{frstr} H. Fritzsch and K.H. Streng, Phys. Lett. B {\bf 72}, 385
(1978). 

\bibitem{hofmann}Ch. Hofmann, G. Soff, A. Sch\"afer and W. Greiner,
Phys. Lett. B {\bf 262}, 210 (1991).

\bibitem{glass}F. Gelis and A. Peshier, Nucl. Phys. A {\bf 697}, 879 (2002).

\bibitem{vjmr} R. Vogt, B.V. Jacak, P.L. McGaughey, and P.V. Ruuskanen, 
Phys. Rev. D {\bf 49}, 3345 (1994).

\bibitem{gmrv} S. Gavin, P.L. McGaughey, P.V. Ruuskanen, and R. Vogt,
Phys. Rev. C {\bf 54}, 2606 (1996).

\bibitem{E691} J.C. Anjos {\it et al.} (E691 Collab.), Phys. Rev. Lett. {\bf
65}, 2503 (1990).

\bibitem{frix2}S. Frixione, in Proc. of {\it EPS01}, 
Budapest, Hungary, 2001, hep-ph/0111368.

\bibitem{herarho}J. Breitweg {\it et al.} (ZEUS Collab.),
Eur. Phys. J C {\bf 2}, 247 (1998).

\bibitem{pbars}A. Trzcinska {\it et al.}, Phys. Rev. Lett. {\bf 87}, 082501
(2001).

\bibitem{e745} T. Kitagaki {\it et al.} (Fermilab E745 Collab.), 
Phys. Lett. B {\bf 214}, 281 (1988).

\bibitem{hpcshad} K.J. Eskola, J. Qiu and J. Czyzewski, private
communication.

\bibitem{eskolanpb} K.J. Eskola, Nucl. Phys. B {\bf 400}, 240 (1993).

\bibitem{ekv} K.J. Eskola, V.J. Kolhinen, and R. Vogt, Nucl.
Phys. A {\bf 696}, 729 (2001).

\bibitem{CDF2}D. Acosta {\it et al.} (CDF Collab.), Phys. Rev. D {\bf 65},
052005 (2002).

\bibitem{matteo}  M. Cacciari and P. Nason, hep-ph/0204025.

\bibitem{refs} R.N. Cahn and J.D. Jackson, Phys. Rev. D {\bf 42}, 3690 (1990);
G. Baur and L.G. Ferreira Filho, Nucl. Phys. A {\bf 518}, 786 (1990); M.
Vidovi\'c, M. Greiner, C. Best, and G. Soff, Phys. Rev. C {\bf 47}, 2308 
(1993); K. Hencken, D. Trautmann and G. Baur, Z. Phys. C {\bf 68}, 473 (1995).

\bibitem{photon01} A. B\"{o}hrer and M. Krawczyk, in Proc. of the
International Conference on the Structure and Interactions of the
Photon, {\it PHOTON 2001}, Ascona, Switzerland, September 2001, to be
published in World Scientific, hep-ph/0203231.

\bibitem{hpcpsi}  R.V. Gavai {\it et al.}, Int. J. Mod. Phys. A {\bf 10} 3043 
(1995); G.A. Schuler and R. Vogt,
Phys. Lett. B {\bf 387}, 181 (1996).

\bibitem{Andreev01} V.P. Andreev, in Proc. of the International
Conference on the Structure and Interactions of the Photon, {\it
PHOTON 2001}, Ascona, Switzerland, September 2001, to be published in
World Scientific.

\bibitem{ohnemus}J. Ohnemus, T.F. Walsh and P.M. Zerwas, Phys. Lett.
B {\bf 328}, 369 (1994).

\bibitem{KLMV} N. Kidonakis, E. Laenen, S. Moch and R. Vogt, Phys. Rev. D {\bf
64}, 114001 (2001). 

\bibitem{RVzphys} R. Vogt, Z. Phys. C {\bf 71}, 475 (1996).

\bibitem{RVinprog} R. Vogt, hep-ph/0207359.

\bibitem{HERA-Bbb} M. Villa (HERA-B Collab.), private communication.

\bibitem{bkn}A.J. Baltz, S.R. Klein and J. Nystrand,
Phys. Rev. Lett. {\bf 89}, 012301 (2002).

\bibitem{ua1}C. Albajar {\it et al.}, Nucl. Phys. B {\bf 335}, 261
(1990).

\bibitem{stara}C. Adler {\it et al.} (STAR Collab.),
Phys. Rev. Lett.  {\bf 87}, 112303 (2001).

\bibitem{pompom}C.G. Roldao and A.A. Natale, Phys. Rev. C {\bf 61}, 064907
(2000). 

\end{references}
\end{document}